\pdfoutput=1
\documentclass{JINST}
\usepackage{graphicx}
\usepackage{amssymb}
\usepackage{makeidx}
\usepackage{amsfonts}
\usepackage{amstext}
\usepackage{amsmath}
\usepackage{amsbsy}
\usepackage{wasysym}
\usepackage{fontenc}
\usepackage{times}
\usepackage{mathptmx}
\usepackage{multirow}
\usepackage{pslatex}
\title{
\vspace{-0.3cm}
VUV-Vis optical characterization of Tetraphenyl-butadiene films on glass and specular reflector substrates
from room to liquid Argon temperature}

\author{
\vspace{-0.3cm}
R. Francini$^a$, R.M.~Montereali$^b$, E.~Nichelatti$^c$, M.A.~Vincenti$^b$, 
N.~Canci$^d$, E.~Segreto$^d$, F.~Cavanna$^e$$^,$$^f$$^,$\thanks{Corresponding author.}~, F.~Di~Pompeo$^e$,
F.~Carbonara$^g$, G.~Fiorillo$^g$, F.~Perfetto$^g$, \\
\llap{$^a$} Dipartimento di Fisica, Universit\`a di Roma Tor Vergata, \\
Via della Ricerca Scientifica 1, 00133 Roma, Italy\\
\llap{$^b$}ENEA C. R. Frascati, UTAPRAD-MNF, Photonics Micro- and Nano-structures Laboratory,\\
Via E. Fermi 45, 00044 Frascati (Roma), Italy\\
\llap{$^c$}ENEA C. R. Casaccia, UTTMAT-OTT, Optical Devices Laboratory, \\
Via Anguillarese 301, 00123 S. Maria di Galeria (Roma), Italy\\
\llap{$^d$}INFN - Laboratori Nazionali del Gran Sasso, \\
s.s. 17 bis km 18+910, 67100 Assergi (L'Aquila), Italy\\
\llap{$^e$}Dipartimento di Scienze Fisiche e Chimiche, Universit\`a dell'Aquila, \\
Via Vetoio, 67100 L'Aquila, Italy\\
\llap{$^f$}Physics Department, Yale University, 
217 Prospect Street, New Haven, CT  06520,  USA\\
\llap{$^g$}Dipartimento di Fisica, Universit\`a "Federico II" and INFN,
 via Cinthia, 80125 Napoli, Italy\\ 

E-mail:\email{flavio.cavanna@aquila.infn.it {\rm or} flavio.cavanna@yale.edu}}

\abstract{
The use of efficient wavelength-shifters from the vacuum-ultraviolet to the photo-sensor's range of sensitivity  is a key feature in detectors for Dark Matter search and neutrino physics based on liquid argon scintillation detection.
Thin film of Tetraphenyl-butadiene (TPB) deposited onto the surface delimiting the active volume of the detector and/or onto the photosensor optical window is the most common solution in current and planned experiments.
Detector design and response can be evaluated and correctly simulated only when the properties of the optical system in use (TPB film + substrate) are fully understood. Characterization of the optical system requires specific, sometimes sophisticated optical methodologies. In this paper the main features of TPB coatings on different, commonly used substrates is reported, as a result of two independent campaigns of measurements at the specialized optical metrology labs of ENEA and University of Tor Vergata. Measured features include TPB emission spectra with lineshape and relative intensity variation recorded as a function of the film thickness and for the first time down to LAr temperature, as well as optical reflectance and transmittance spectra of the TPB coated substrates in the wavelength range of the TPB emission.
}
\preprint{arXiv:1304.6117 [physics.ins-det]}

\begin{document}

\section{Introduction}
\label{Introduction}
One of the most intriguing topics in frontier physics today is the nature of Dark Matter (DM) in the Universe. Well motivated candidate are the Weakly Interacting Massive Particles (WIMPs) arising in many of the potential extensions of the Standard Model of particle physics. WIMPs could be detected directly by their collisions with target nuclei in underground experiments. However, since the predicted signal rates are much lower than one interaction per kg of target material per day, large detector masses, high sensitivity to low energy deposits and ultra-low backgrounds are necessary ingredients of any experiment aiming to discover WIMPs. Recent years have seen an increasing number of experiments using noble liquids like Argon or Xenon as materials for detecting WIMPs. These noble liquids have very high scintillation and charge yields in response to ionization events. Detectors collecting both signals (two-phase detectors) or only the scintillation light (single phase detectors) are currently in use and are likely scalable to larger masses (several ton) in the near future.
In liquid argon (LAr) the time dependence of the scintillation light emission is significantly different for heavy ionizing particles, such as nuclear recoils induced by possible WIMP interactions, when compared to light ionizers like electron recoils induced by gamma (overwhelming) background from natural radioactivity.  Background rejection and signal energy threshold greatly improve with scintillation light collection efficiency. Light collection efficiency is therefore the primary requirement in liquid argon detectors for Dark Matter search. 

The light emission in LAr is characterized by an intrinsic band spectrum around 128 nm in the vacuum ultra-violet spectral region (VUV). Direct VUV detection, although possible, can be problematic due to the inability to transmit or guide VUV radiation to a photo-sensor; the use of efficient wavelength-shifters from the VUV to the photo-sensor's range of sensitivity is the most common solution to overcome this difficulty. Fluorescent materials are often used in the detection of extreme ultraviolet radiation. These materials emit visible light upon the absorption of VUV radiation. Tetraphenyl-butadiene (TPB) is the most commonly used wavelength shifter (organic) compound, especially suitable for applications requiring high sensitivity in the VUV region. TPB can be evaporated or deposited by dipping as thin films, yielding translucent, adherent, durable coating. \\
A typical VUV detection system for DM search consists of an array of photosensitive devices (e.g. photomultipliers) viewing a LAr volume target (typically immersed in a uniform electric field) whose inner surfaces are TPB coated to obtain the highest scintillation light collection efficiency. Similar schemes are also adopted in present and forthcoming massive LAr neutrino detectors, though with reduced 
photo-cathodic coverage and only partial TPB coating of the inner surfaces.\\
The optical properties of a TPB film depend (even significantly) on the film thickness and the deposition process but also on the type of substrate [e.g. polymer plastic layer, or Polytetrafluoroethylene (PTFE) or acrylic  plastic tiles that delimit the LAr volume or the glass of the photomultiplier tube (PMT) windows]. Operating in a cryogenic environment (87 K for Argon in liquid phase) introduces an additional dependence on temperature. 
Detector design and response can be evaluated and correctly simulated only when the properties of the optical system in use (TPB film + substrate) are fully understood. Characterization of the optical system requires specific, sometime sophisticated optical methodologies. In this paper the main features of TPB coatings on different substrates are reported, as a result of two independent campaigns of measurements at the specialized optical metrology labs of ENEA and University of Tor Vergata. Measured optical features include TPB photoemission spectra with lineshape and relative intensity variation as a function of temperature (down to LAr temperature) and film thickness, as well as optical reflectance and transmittance of the TPB coated substrates in the wavelength range of the TPB emission.

\subsection{Down-conversion of LAr scintillation light}
\label{sec:downconv_vuv}

For Ar in liquid phase (boiling point T=87.3~K), the scintillation emission spectrum \cite{morikawa}
lies in the VUV range, with peak wavelength at
\begin{equation}
\lambda~\simeq~~ 128~{\rm nm}; 
\end{equation}
which corresponds to the 9.69 eV peak of a Gaussian energy band,  0.45 eV of full width (FWHM).

Glass exhibit null transmittance for VUV photons and light has to be shifted to longer wavelengths if PMTs\footnote{Only $MgF_2$ windows provides non-negligible transmittance at short (VUV) wavelengths, however the use of this material is unpractical due to its fragility at low temperature and to the high cost.}  are being used as photo-detectors. This can be accomplished by using {\itshape wavelength shifter}  aromatic hydrocarbons such as {\itshape Tetraphenyl-butadiene}. \\
TPB (1,1,4,4-Tetraphenyl-1,3-butadiene, C$_{28}$H$_{22}$) is a long-recognized best suited material for wavelength downconversion from VUV to blue-visible \cite{burton}. The emission spectrum of TPB is peaked at around 440~nm and extends from 390 to 520~nm, where the transmittance of the glass window and the photocathode quantum efficiency of the PMTs are sufficiently high \cite{burton, samson}. 
TPB fluorescence efficiency in the excitation wavelenght range from 90 nm to 250 nm was reported to be very high, largely exceeding Sodium Salicylate, normally used  as standard reference material \cite{brunet,hoang-mai}. Recent measurements in the VUV range \cite{gehman} report total integrated fluorescence efficiency values  greater than unity. According to the material energy levels, the interaction of a (V)UV photon can induce in some case the production of more than one visible photon. 

Advancement of TPB coating techniques quickly developed after the initial work of McKinsey et al. \cite{mckinsey_tpb1}.
To efficiently provide LAr based detectors with a high light detection capabilities  and to make homogeneous the detector response, all the internal surfaces delimiting the LAr sensitive volume have to be made of highly reflecting layers coated by a TPB film.  Scintillation VUV photons from energy deposition in the LAr volume propagate inside the LAr volume, then are wavelength-shifted into visible photons  when hitting the TPB film on the surface boundaries and finally the visible photons are reflected (several times) from the surfaces beneath, up to collection from the active (photo-cathodic) area of the PMT. 
This is possible because the TPB film, in first approximation, does not absorb the visible photons. Collection efficiencies of shifted photons exceeding 60$\%$ can be reached with a  moderate photo-cathodic coverage ($\sim$15$\%$) if the reflectivity of the internal surface is high enough (above $90\%$) \cite{secret}.
 In addition, the PMT windows can also be coated with a very thin TPB layer whose thickness and deposition method are optimized to maximize the transmission of the blue shifted photons and to (at least partially) convert the VUV light emission component that directly hits the photo-sensitive area of the detector.

The reflector surface delimiting the LAr volume can be either diffusive, e.g. made by PTFE tiles or tissue, or specular, e.g. by polymeric multi-layer plastic totally dielectric mirrors foils \cite{janecek}. TPB film on these surfaces, typically in the 150 to 1500 $\mu$g/cm$^2$ density range, is obtained by deposition with vacuum evaporation technique. TPB coating of the PMT glass window (50 $\mu$g/cm$^2$ or less) can also be obtained by evaporation, or embedded in a polystyrene matrix (5\% to 25\% weight fraction, deposited by dip-coating in toluene solvent) for a more durable and adherent film deposition. Reflectance of the boundary surfaces after TPB film deposition and transmittance of PMT glass with TPB-polystyrene coating are fundamental detector parameters to be determined.  

\section{Sample preparation}
\label{sec:sample_prep}
The following optical systems have been investigated and reported in this paper:
\begin{itemize}
\item TPB film on glass substrate
\item TPB film on specular reflector substrate
\item TPB in polystyrene matrix on glass substrate
\end{itemize}
Samples were produced by thermal evaporation at different thicknesses or by dip-coating with solution at 
different TPB concentrations. Only best scintillation grade TPB (>99.0\% pure) was used.\\
\begin{table}[!htbp]
\centering
\caption{{\scriptsize \sf Investigated samples obtained by thermal evaporation of TPB on different substrates. The layer density is evaluated by precise measurement of the weight difference of the sample before and after TPB evaporation.}}
\vspace{0.1cm}
\begin{tabular}{|c|c|c|c|}
	\hline
Substrate type & TPB layer density  & Substrate type & TPB layer density \\ 
                           &  [$\mu$g/cm$^2$]   &                            & [$\mu$g/cm$^2$]   \\ \hline\hline
\multirow{12}{*}{Glass }  & 0 & \multirow{12}{*}{Polymeric Reflector} &0      \\ 
                                                          & 165 &      &   50      \\ 
                                                          & 200 &      & 75      \\ 
                                              		  & 350 &      & 175      \\ 
		                                       & 390 &      & 370      \\  
                                               	  & 725 &      & 475      \\  
                                              		  & 745 &	     & 550     \\ 
	                                                 & 900 &       & 600     \\	                                                 
                                              		  & 1160 &     &  770      \\                                                 
	                                                 & 1175 &     & 810      \\ 
                                              		  & 1450 &     & 1045    \\  
                                              		  & 1800 &	     &  1375    \\ 
                                          	           & 1875 &      & 1500     \\
	  \hline\hline                                              
\end{tabular}
\label{evap-samples}
\end{table}

\begin{figure}[ht]
\begin{center}
\includegraphics[width=15.cm]{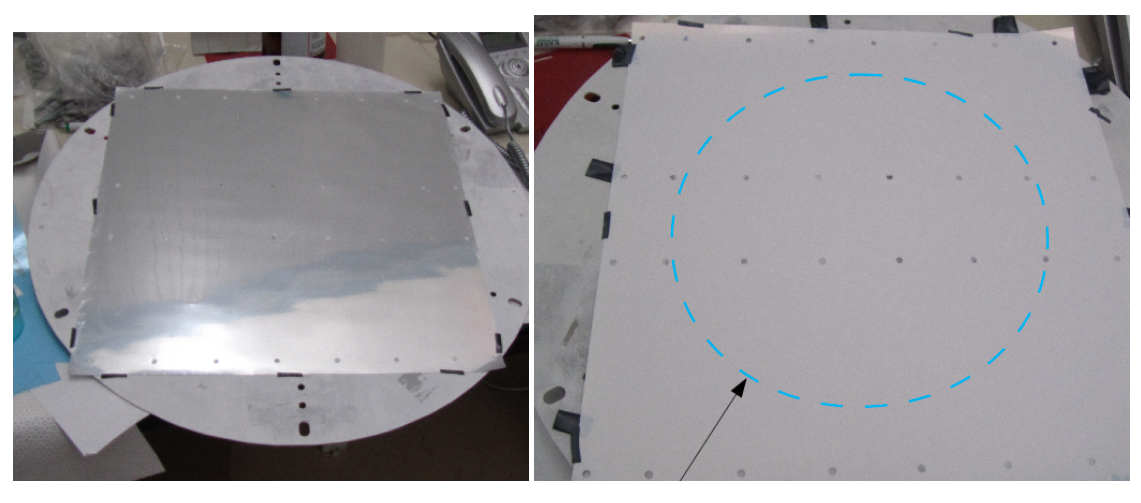}
\caption{{\scriptsize \sf Picture of a reflector layer before and after TPB evaporation. The layer (originally mirror-like) appears bright white due to light diffusion properties of the TPB film deposited on the mirror surface.}}
\label{fig:evap_layers}
\end{center}
\end{figure}

The reflector substrate is a polymeric multi-layer plastic mirror totally dielectric 
(Vikuiti\texttrademark ~ESR by  3M$\textsuperscript{\textregistered}$ \cite{vikuiti}, formerly known as VM2000\texttrademark) with very high specular reflectivity ($>$ 98\%) \cite{vikuiti,janecek} constant across the visible spectrum. 
This is a film like product, flexible and lightweight. The material is metal free, non-corrosive and non-conducting. It is  low-shrinkage and thermally stable down to low temperatures. 
The TPB film on the reflector substrate is obtained by deposition with vacuum evaporation technique. \\
A custom evaporation system for the TPB deposition on reflector or on glass was designed and realized (INFN Pavia) in 2007, in the framework of the WArP experiment for DM search at the Gran Sasso Laboratory (Italy).
The evaporation system is based on a vacuum chamber, equipped with an evaporation crucible (Knudsen cell) optimized for melting and evaporation of organic compounds.
The Knudsen cell utilizes the principle of molecular effusion. The material to be deposited is heated to provide a suitable vapor pressure in an isothermal enclosure. Molecular effusion from an orifice on the top side of the cell gives rise to a cosine intensity distribution. The deposition rate is extremely stable being determined by the temperature of the furnace which is accurately controlled with a PID controller (Proportional Integral Derivative). The copper furnace of the cell is designed as a removable cartridge (mounted on the bottom side of the vacuum chamber) which contains the crucible, heater element and heat shields. The heater filament is a tantalum foil element which is isolated with PBN shields (Pyrolitic Boron Nitride). The temperature of the furnace is monitored by a thermocouple.  \\
The system (vacuum system, handling, etc.) is designed to perform an evaporation cycle in about  two hours. This system is operated at the Gran Sasso Lab (external laboratory) where a dedicated (clean) area has been made available. After preparation, TPB samples (as listed in Tab.\ref{evap-samples}) are sealed in nylon bags filled with inert gas (pure Argon), and stored in dark up to the experimental test.
In Fig.\ref{fig:evap_layers}, one can see a reflector layer before and after TPB evaporation.

Films formed by TPB in polystyrene (PS) matrix were deposited on clean glass slides under ambient conditions. First, solutions were made in toluene by mixing pre-measured quantities of TPB and pure PS in grains. Solutions were mixed with an heated magnetic stirrer at a temperature of 40$^o$C for 30 min in order to make them perfectly homogeneous. 
The glass substrate is dipped in the solution, withdrawn and then left to dry in air. A thin adhesive transparent film is left on the glass surface. Also in this case, after preparation samples (as listed in Tab.\ref{dip-coat-samples}) are sealed in nylon bags in inert gas atmosphere.

\begin{table}[!htbp]
\centering
\caption{{\scriptsize \sf Investigated samples obtained by dip-coating of TPB in polystyrene on glass substrates.}}
\vspace{0.1cm}
\begin{tabular}{|c|c|}
	\hline
Substrate type & TPB in polystyrene matrix \\
                           & Weight fraction [\%] \\ \hline\hline
\multirow{4}{*}{Glass }  & 1 \\ 
						  &10 \\ 
						  & 20 \\  
						  & 40 \\ \hline\hline
\end{tabular}
\label{dip-coat-samples}
\end{table}

\section{PhotoLuminescence and PhotoExcitation}
\label{sec:PL-PLE}

The efficiency of the wavelength shifting mechanism depends on
the wavelength of the interacting photons and on the properties of the fluorescent material.

The wavelength-shifting efficiency can be measured at a reference wavelength $\lambda_{emi}$ of the emission spectrum, as a function of the  wavelength of the incident light (source). 
In this case, the spectral behavior of the conversion efficiency (e.g. from near infrared - NIR - to VUV) is called {\it PhotoExcitation} (PE), or {\it Excitation PhotoLuminescence spectrum},  and represents the spectral emission characteristic of the wavelength-shifting (WLS) material as a function of the absorbed light wavelength.

The wavelength distribution of the converted light conversely can be obtained under pumping at a reference excitation wavelength 
$\lambda_{exc}$ in the absorption spectral range and recording the amplitude as a function of the wavelength of the emitted light.
The spectra obtained by exciting the sample at a given incident wavelengths are called {\it PhotoLuminescence {\rm (PL)} spectra}  and represent the WLS emission spectra.
PL spectra may change in shape and amplitude with the incident wavelength. 

\subsection{Experimental set-up}
\label{sec:exp_set-up}
Two sets of TPB PL and PE measurements with different samples have been performed, the first set in UV-Vis spectral range at room temperature (RT)
and the second one extending to the VUV range and exploring - for the first time to our knowledge - the effect of temperature (from 
RT to LAr T) on the WLS properties.

At optical wavelengths in UV-Vis spectral range, a HORIBA Jobin Yvon Fluorolog$\textsuperscript{\textregistered}$-3 mod. FL-11 spectrofluorimeter was utilized in a front-face detecting geometry to measure at room temperature the PL and PE spectra of several TPB samples of different thickness. The spectrofluorimeter is equipped with a 450 W xenon excitation lamp, which provides a continuous spectral distribution in a range from 200 nm to 600 nm, two Czerny-Turner single grating (1200 grooves/mm) monochromators, one for excitation, blazed at 330 nm, and the other one for emission, blazed at 500 nm. The Fluorolog$\textsuperscript{\textregistered}$ is also provided with a photodiode for intensity excitation correction within 240-1000 nm and a Hamamatsu R928P photomultiplier operating in photon-counting mode as emission detector.

VUV PhotoExcitation and PhotoLuminescence measurements were performed using as excitation source a 30 Watt deuterium lamp (McPherson\texttrademark Model 632), equipped with a magnesium fluoride (MgF$_2$) window delivering a continuous spectrum ranging from 115 nm to 300 nm. The whole optical path from the MgF$_2$~ window down to the sample was kept in vacuum. The light from the lamp was collected and focused on the entrance slit of a 20 cm focal length vacuum grating spectrometer (McPherson\texttrademark Model 234/302) operating with a 1200 g/mm aberration corrected concave grating. Light from the monochromator exit slit was focused inside the sample chamber by a collimator mirror setup, and impinged on the sample at 45$^o$, exciting an area of a few square millimeters, depending on the slit width. Luminescence from the sample was collected  at 90$^o$ with respect to the excitation beam, through a quartz window, after which the measurement setup was in air. It consisted of a collimating lens arrangement, a 30 cm focal length grating monochromator (ARC Model SpectraPro$\textsuperscript{\textregistered}$-300i), equipped with a 1200 g/mm plane grating blazed at 500 nm. The signal was detected by a low dark counts multialkali photomultiplier, operating in single-photon counting regime. For the measurements in the temperature range 87~K - 300~K the samples were mounted on the cold finger of a closed-cycle helium cryostat, while keeping the same excitation and detection geometry.

\subsection{TPB evaporated on glass substrate}
\begin{figure}[!htbp]
\centering
\includegraphics[width=9.5cm]{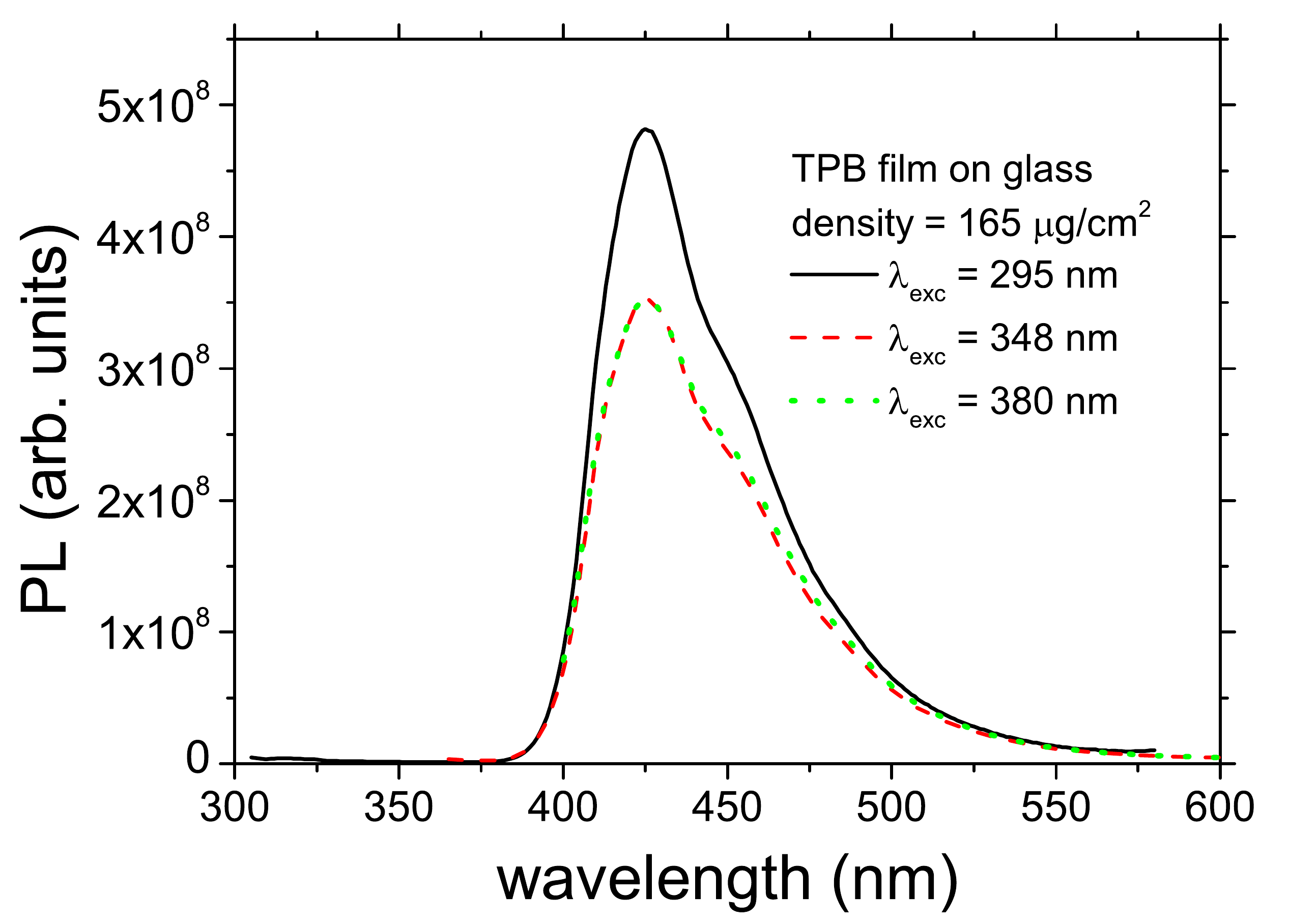}
\caption{{\scriptsize \sf RT photoluminescence spectra of TPB sample evaporated on glass substrate (layer density 165 $\mu$g/cm$^2$) excited at three wavelengths. } }
\label{fig:R1} 
\end{figure}

PL spectra have been recorded from TPB samples deposited on glass substrate at various thicknesses,  with excitation wavelength in the UV-Vis range as well as in the VUV range at 128 nm. \\
As example, Fig.\ref{fig:R1} shows  the PL spectra measured at RT of the TPB sample evaporated on glass with 165 $\mu$g/cm$^2$ density excited at $\lambda_{exc}$=~295 nm, 348 nm and 380 nm.
The spectra are corrected for the excitation-lamp intensity at these wavelengths.
The emission band extends from 390 nm to 550 nm and is peaked at about 425 nm;
 its lineshape is the same at all excitation wavelengths indicating that from the excited states of TPB, non-radiative relaxation brings the molecule to same lower excited state from which the 425 nm emission originates. A shoulder appears at about 450 nm followed by a long tail toward longer wavelengths. These structures can be attributed to a poorly resolved vibronic sequence.
\begin{figure}[!htbp]
\centering
\includegraphics[width=9.5cm]{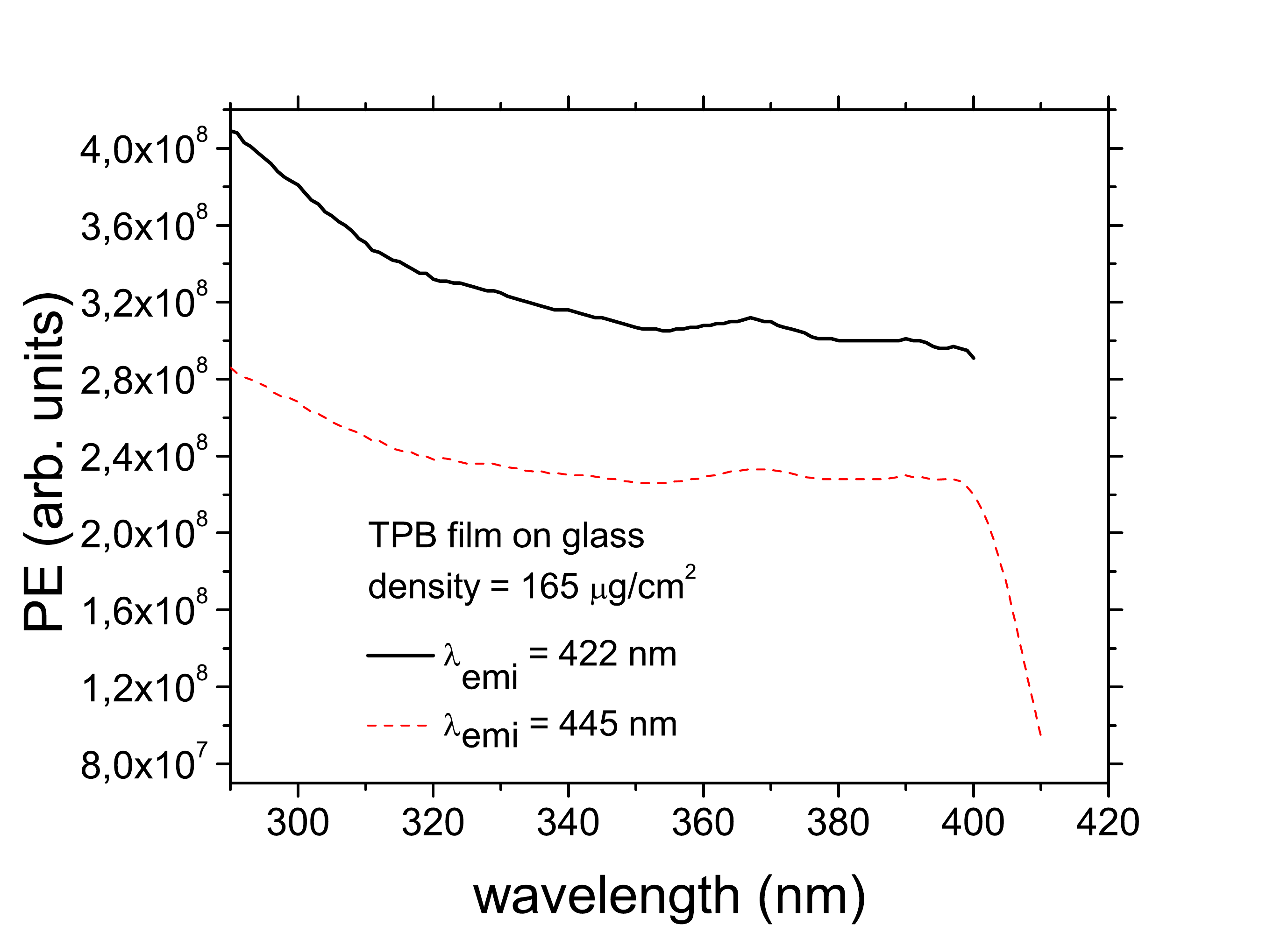}
\caption{{\scriptsize \sf RT photoexcitation spectra of TPB sample evaporated on glass substrate (layer density 165 $\mu$g/cm$^2$) at two different emission wavelengths.} }
\label{fig:R2} 
\end{figure}
For the same sample, we report in Fig.\ref{fig:R2} the PE spectra measured at two emission wavelengths $\lambda_{emi}$ =~422 nm and 445 nm within the luminescence band. The excitation spectrum does not reveal any significant structure in the range 290 nm - 400 nm.
One can see the onset of TPB excitation around 410 nm, indicating a significant overlap in the $\sim$(390-410) nm range between absorption and emission spectra. 
This overlap can affect the high energy tail of TPB emission, inducing an apparent red shift of the emission band as well as a decrease of the emission band full width, depending on the thickness of the evaporated TPB layer. \\
\begin{figure}[!htbp]
\centering
\includegraphics[width=10.cm]{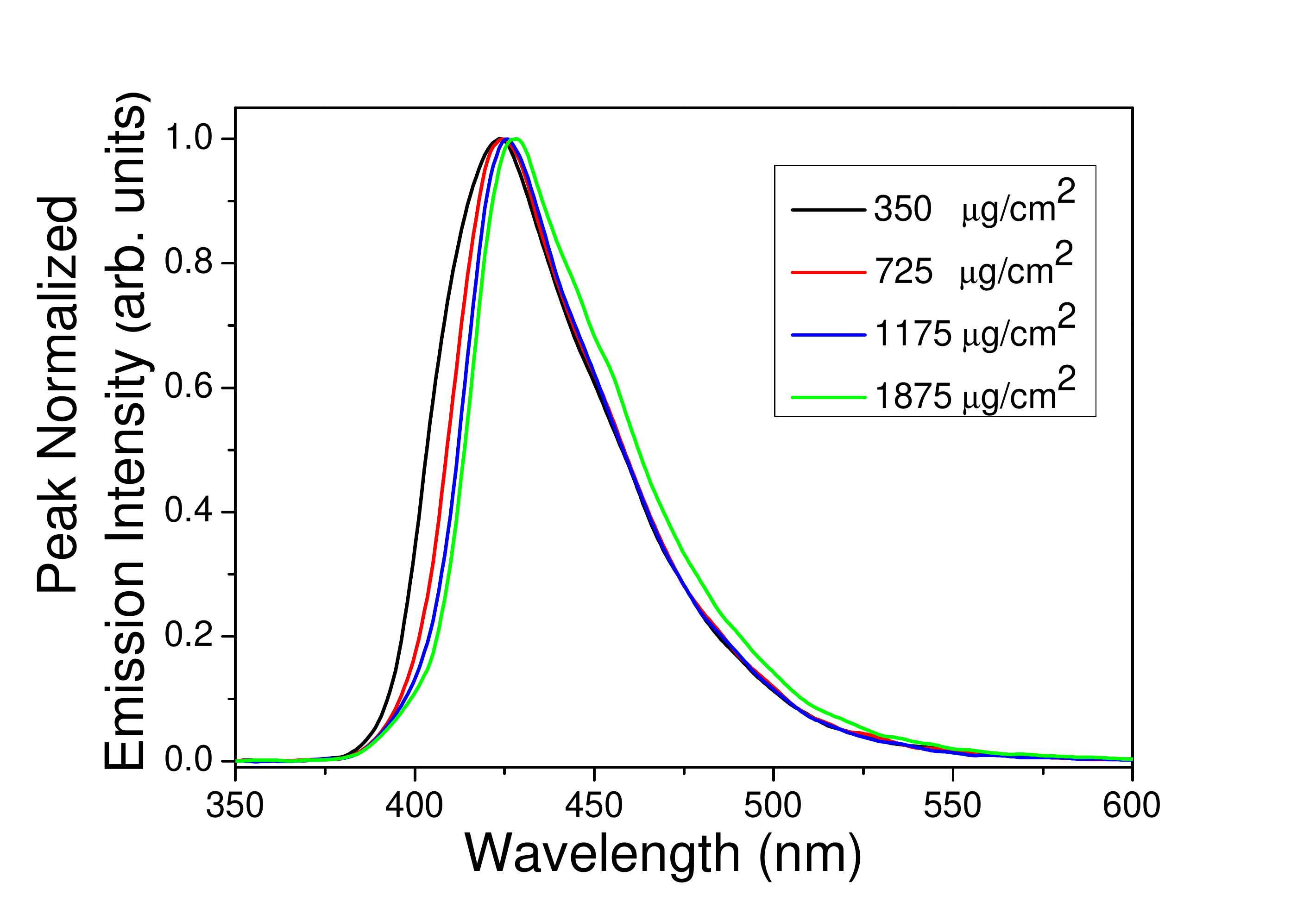}
\caption{{\scriptsize \sf Normalized RT photoluminescence spectra of TPB samples evaporated on glass substrates recorded at four different layer densities, excited at 128 nm.} }
\label{fig:R3} 
\end{figure}
\begin{figure}[!htbp]
\centering
\includegraphics[width=10.cm]{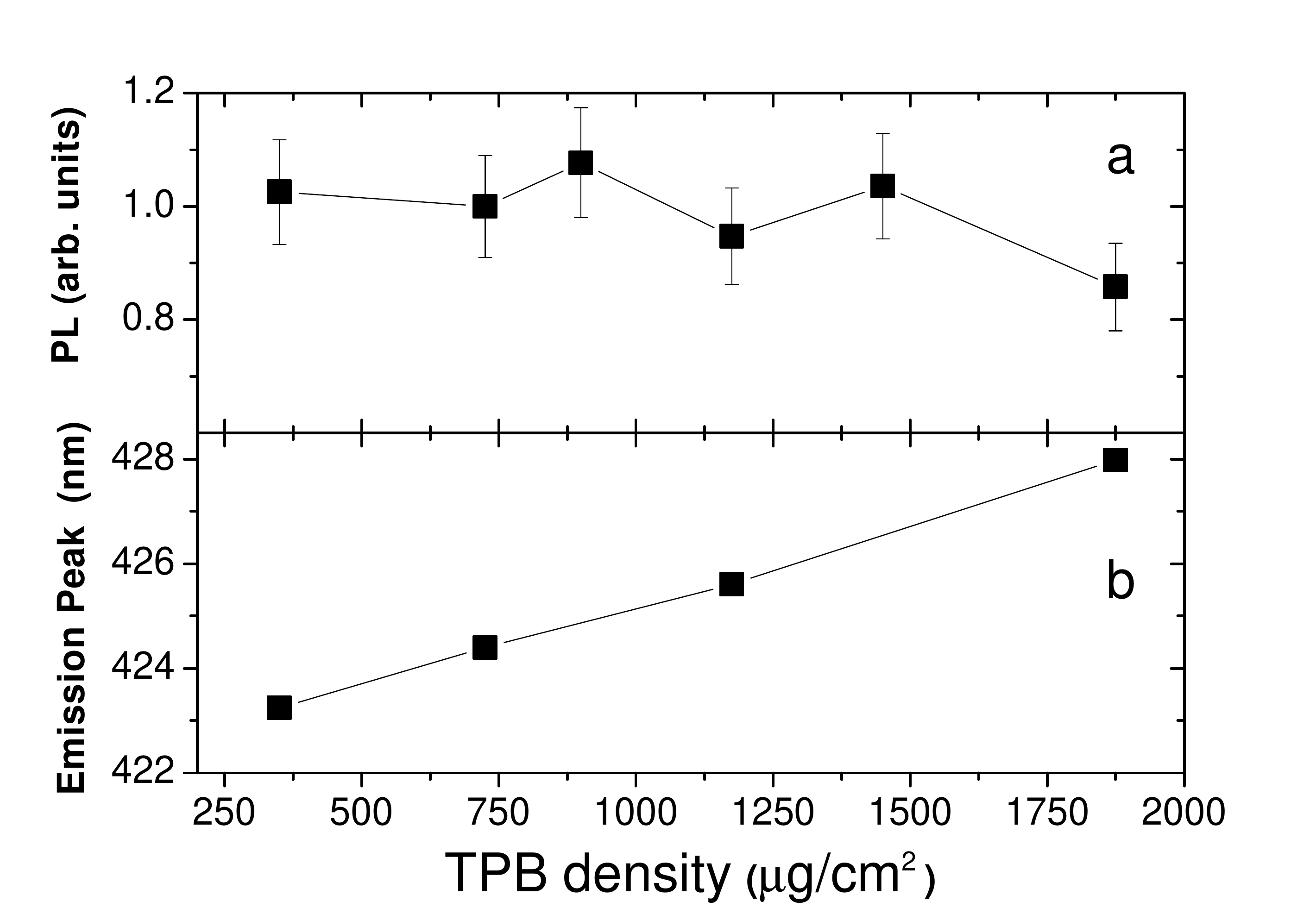}
\caption{{\scriptsize \sf Integrated relative photoluminescence intensities (a) and peak wavelength position (b) as a function of the TPB layer density evaporated on glass substrate.} }
\label{fig:R4} 
\end{figure}
Such an effect can be found in the PL spectra shown in Fig.\ref{fig:R3} of four samples with different TPB layer thickness, measured at RT. The PL spectra in Fig.\ref{fig:R3} are excited in the VUV region at $\lambda_{exc}$=~128 nm, and one should first notice that the emission bands do not significantly differ from those shown in Fig.\ref{fig:R1}. The integrated relative PL intensity  as a function of the layer density is plotted in Fig.\ref{fig:R4} [Top] and it is found to be constant, within experimental uncertainty, from 350 $\mu$g/cm$^2$ up to 1450  $\mu$g/cm$^2$ followed by a significant decrease at 1875  $\mu$g/cm$^2$.

As the thickness of the TPB layer increases, the emission peak and the high energy tail shift toward the red as an effect of re-absorption. Evaporated TPB samples are finely grained films which act as good scatterers for the emitted radiation, a fraction of which can therefore undergo several re-absorption and re-emission processes before escaping from the sample surface. The resulting red shift of the emission band peak is shown in Fig.\ref{fig:R4}-b.  

\subsection{TPB evaporated on specular reflector substrate}
\label{sec:TPBonVIK}
As for the samples evaporated on glass, TPB films evaporated on polymeric mirror substrates (VM2000\texttrademark~ or Vikuiti\texttrademark) have been optically characterized in the near UV and in the VUV ranges.

\begin{figure}[!htbp]
\centering
\includegraphics[width=7.5cm]{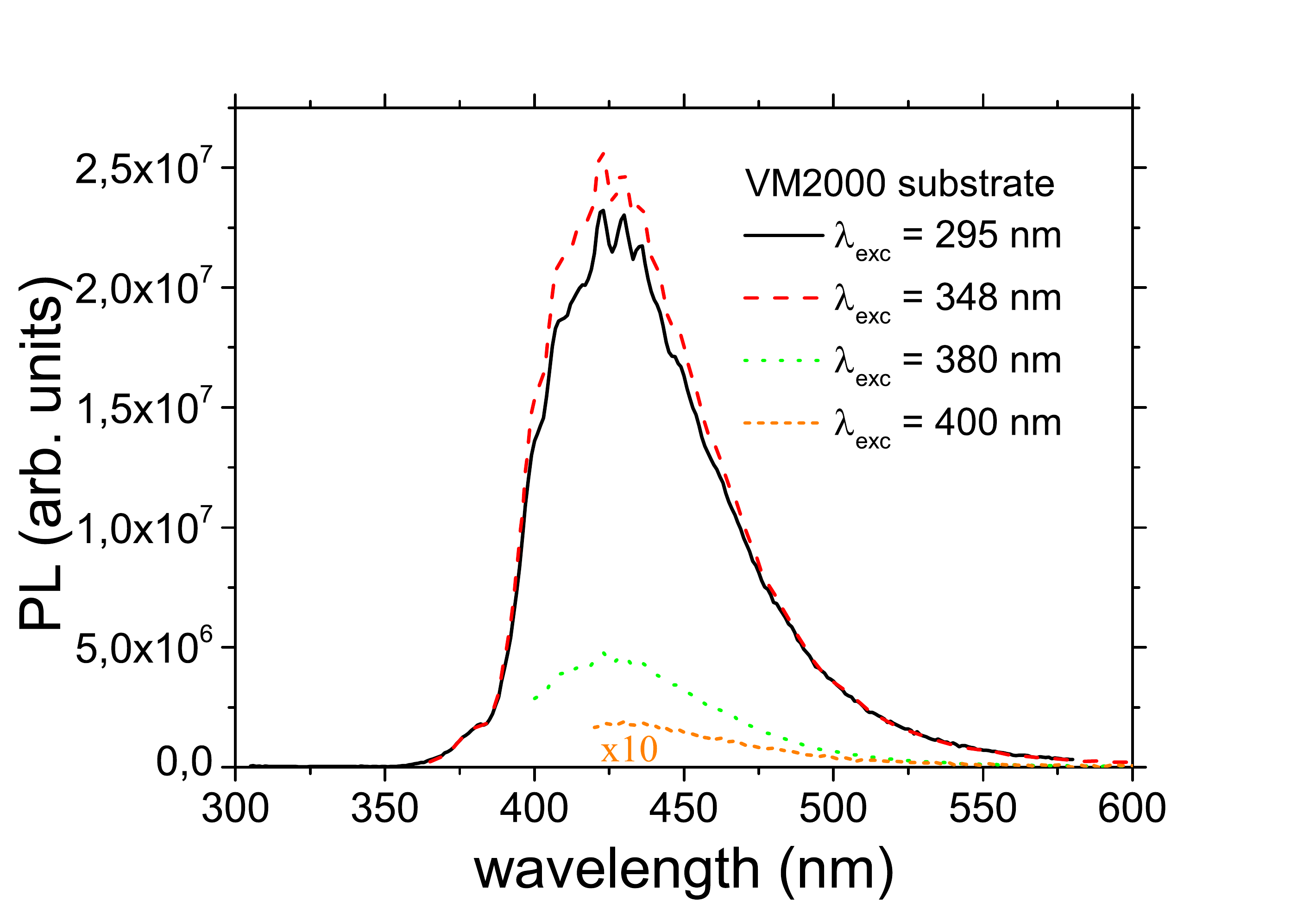}
\includegraphics[width=7.3cm]{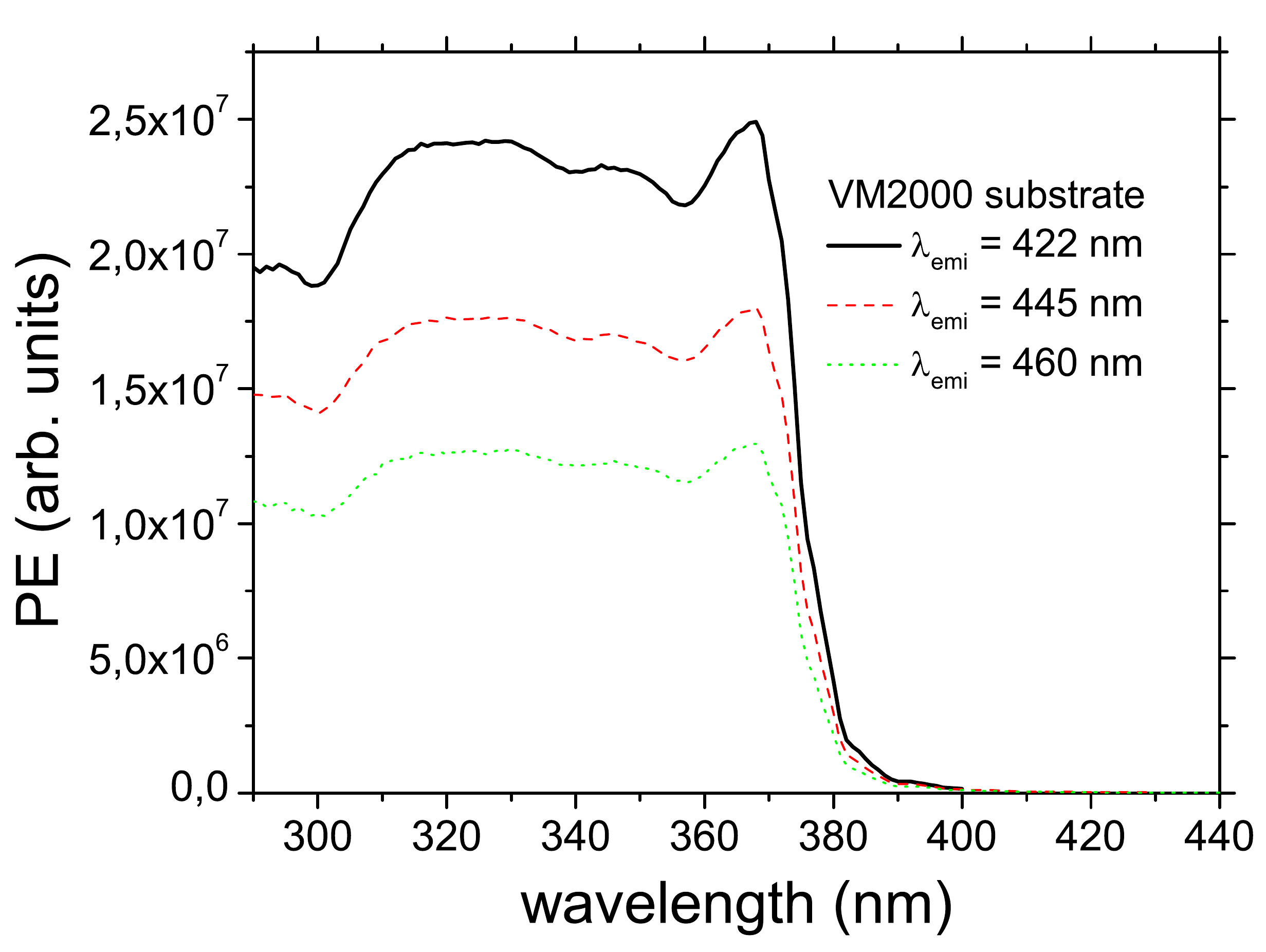}
\caption{{\scriptsize \sf Bare polymeric substrate: [Left] RT photoluminescence spectra excited at different UV-Vis wavelengths and 
[Right] RT photoexcitation spectra at three different emission wavelenghts.} }
\label{fig:PL_PE_VM2000} 
\end{figure}

The photoemission properties of the bare polymeric reflector substrate were first measured by recording its PL and PE spectra shown in Fig.\ref{fig:PL_PE_VM2000}. PL and PE spectra were then acquired in the same experimental conditions from TPB samples evaporated at several layer densities on polymeric reflectors (Tab.\ref{evap-samples}). 

\begin{figure}[!htbp]
\centering
\includegraphics[width=7.5cm]{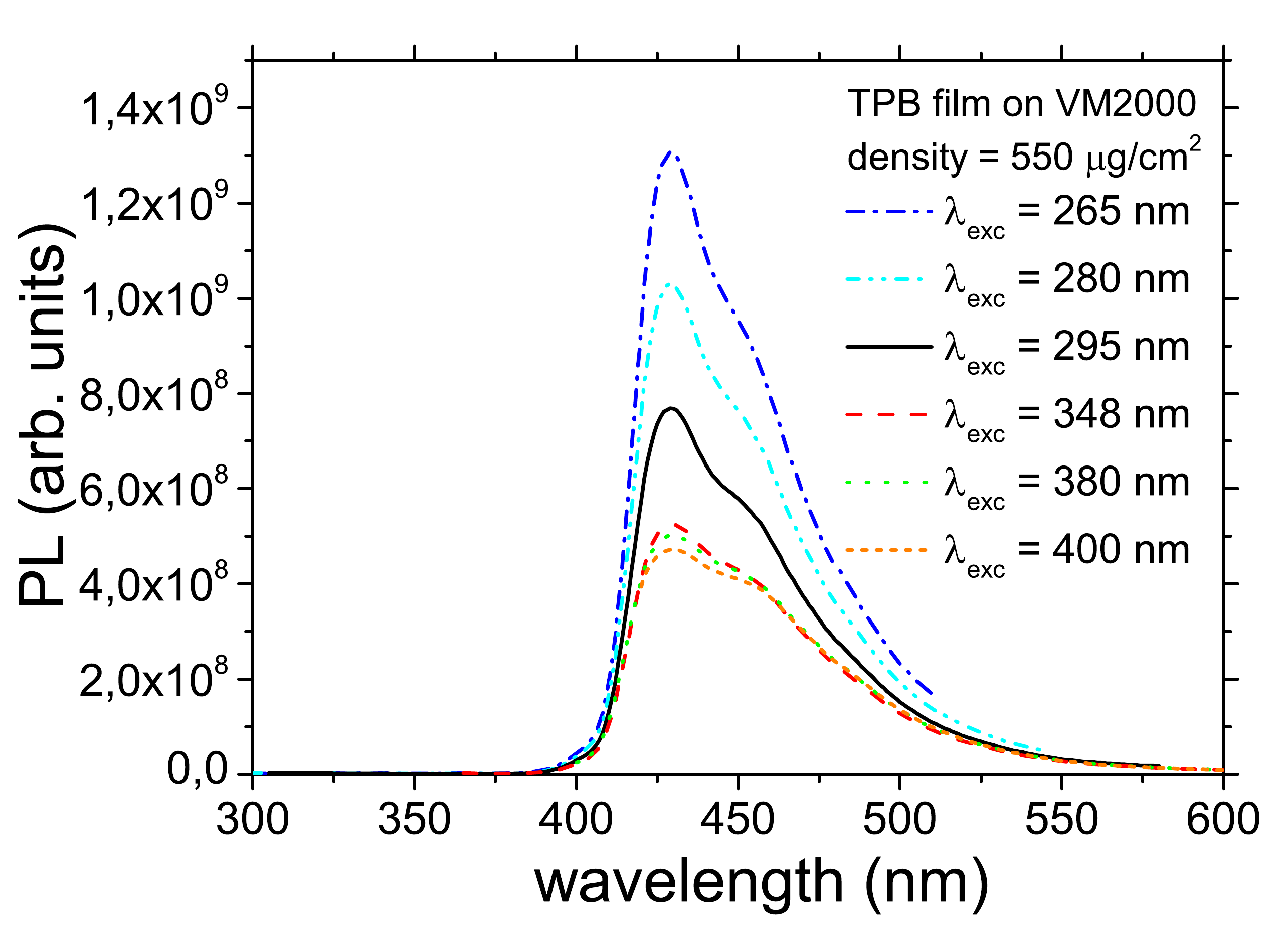}
\includegraphics[width=7.3cm]{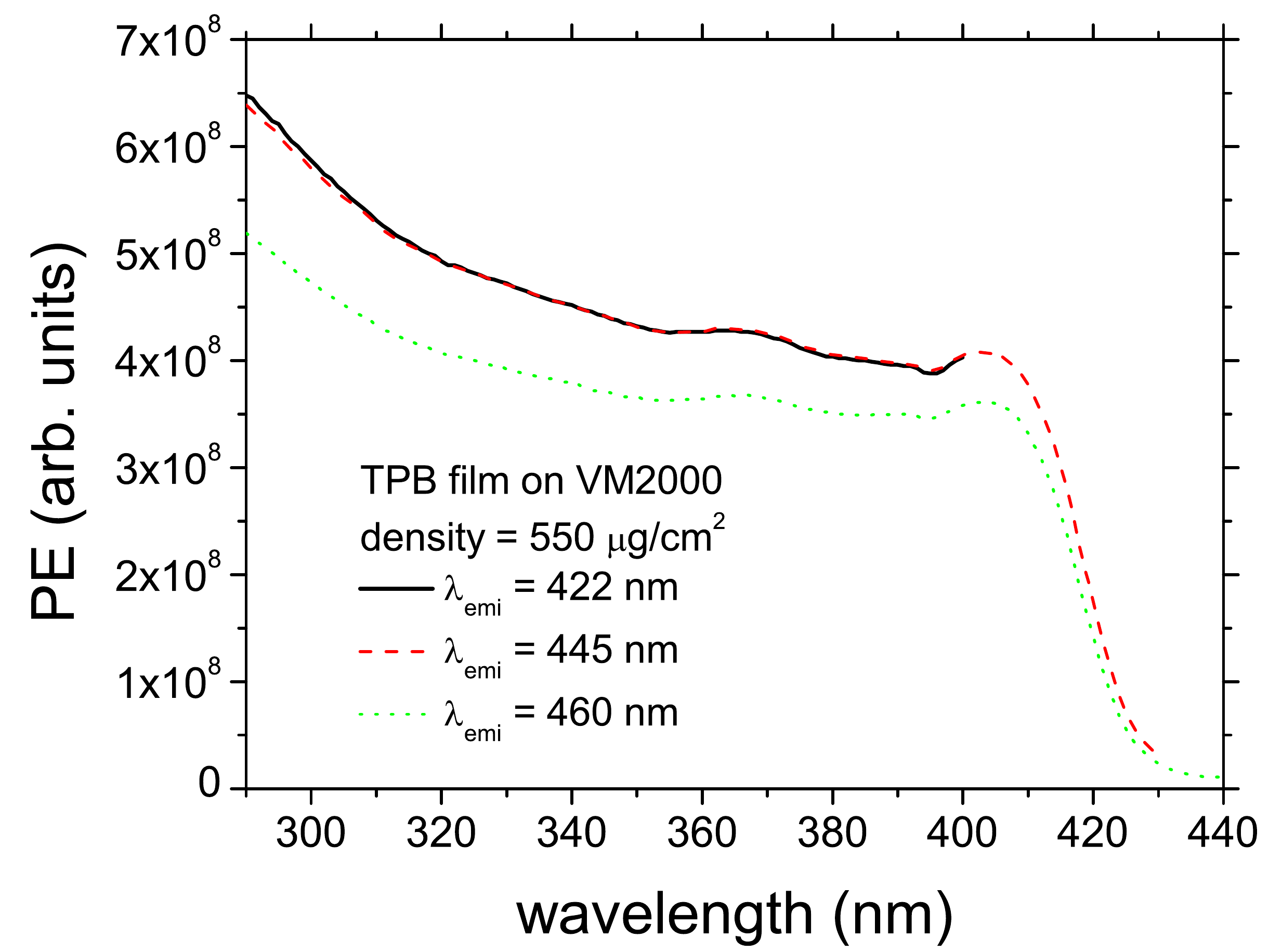}
\caption{{\scriptsize \sf TPB sample evaporated on polymeric reflector substrate (layer density 550 $\mu$g/cm$^2$): [Left] RT photoluminescence spectra excited at different UV-Vis wavelengths and 
[Right] RT photoexcitation spectra at three different emission wavelengths.} }
\label{fig:R5} 
\end{figure}

As example, the PL spectra of a 550 $\mu$g/cm$^2$ TPB film on reflector (VM2000\texttrademark)~ recorded at RT for several UV excitation wavelengths are reported in Fig.\ref{fig:R5} [Left]. The lineshape is very similar to that found in the case of the glass substrate but the emission peak is red-shifted to 430 nm, whereas those structures ascribed to vibronic replicas do not appear to be shifted. The polymeric substrate behaves like a mirror in the spectral region where the TPB emission is centered \cite{vikuiti}. 
This fact could explain the peak shift in terms of a more pronounced re-absorption effect, taking into account that a portion of the emitted light, after reaching the substrate is reflected back and is eventually collected by the detection setup. This portion of luminescence was lost in the case of the glass substrate, escaping from the back surface of the latter.
 A longer path of the emitted light through the TPB film, as discussed previously, could result in a enhanced red shift of the PL peak. 
For the same sample, the PE spectra measured at reference wavelengths $\lambda_{emi}$ =~422 nm, 445 nm  and 460 nm in the luminescence band is reported in Fig.\ref{fig:R5} [Right]. The onset of TPB excitation is around 420 nm, with a significant overlap in the $\sim$(400-420) nm range between absorption and emission spectra. The excitation spectra do not reveal any structure in the range 290 nm - 400 nm. From the comparison of the signal intensities in Fig.\ref{fig:PL_PE_VM2000} and Fig.\ref{fig:R5}, the photoluminescence emitted by the bare polymeric substrate can be considered negligible as it is more than an order of magnitude lower than that emitted by the TPB films.

\begin{figure}[!htbp]
\centering
\includegraphics[width=11.0cm]{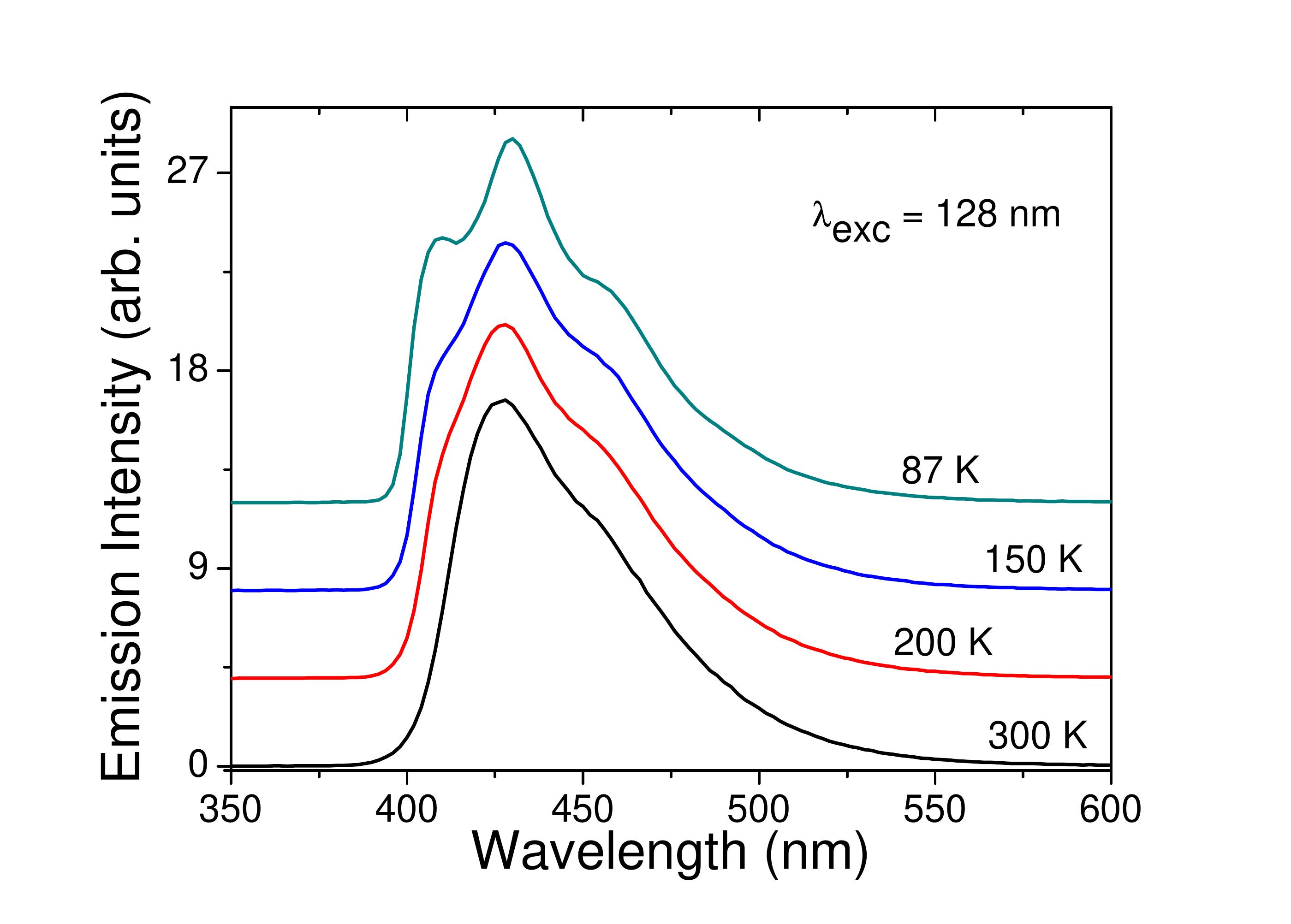}
\caption{{\scriptsize \sf Photoluminescence spectra of TPB evaporated on polymeric reflector substrate (layer density 175 $\mu$g/cm$^2$) excited at 128 nm, as a function of temperature. The emission spectra are vertically shifted for clarity.} }
\label{fig:R6} 
\end{figure}

TPB films evaporated on polymeric reflector substrates (Vikuiti\texttrademark)~ have been characterized by VUV excitation (128 nm) as a function of temperature, in the range 87 K - 300 K, and of the film density. For the 175 $\mu$g/cm$^2$ sample we show in Fig.\ref{fig:R6} the temperature dependence of the PL spectra excited at 128 nm. 
The unresolved vibronic structures at 300 K become clearly visible as the temperature is decreased, and at liquid Argon temperature (87 K) we can observe at least four distinct structures. One can notice that the peak position of these structures do not change with temperature, in agreement with the picture of weakly interacting TPB molecules. For the same reason, by lowering the temperature we observe a marked decrease of the linewidth of each structure, indicating that the homogeneous contribution to the linewidth, arising from the internal degrees of freedom of the TPB molecule, is at least comparable to the inhomogeneous contribution, the latter being a measure of the strength of the interaction of TPB with the surrounding environment (in the case of our evaporated films, the environment is made of other TPB molecules, the arrangement of which can vary from site to site, producing a local disorder).
\begin{figure}[!htbp]
\centering
\includegraphics[width=10.cm]{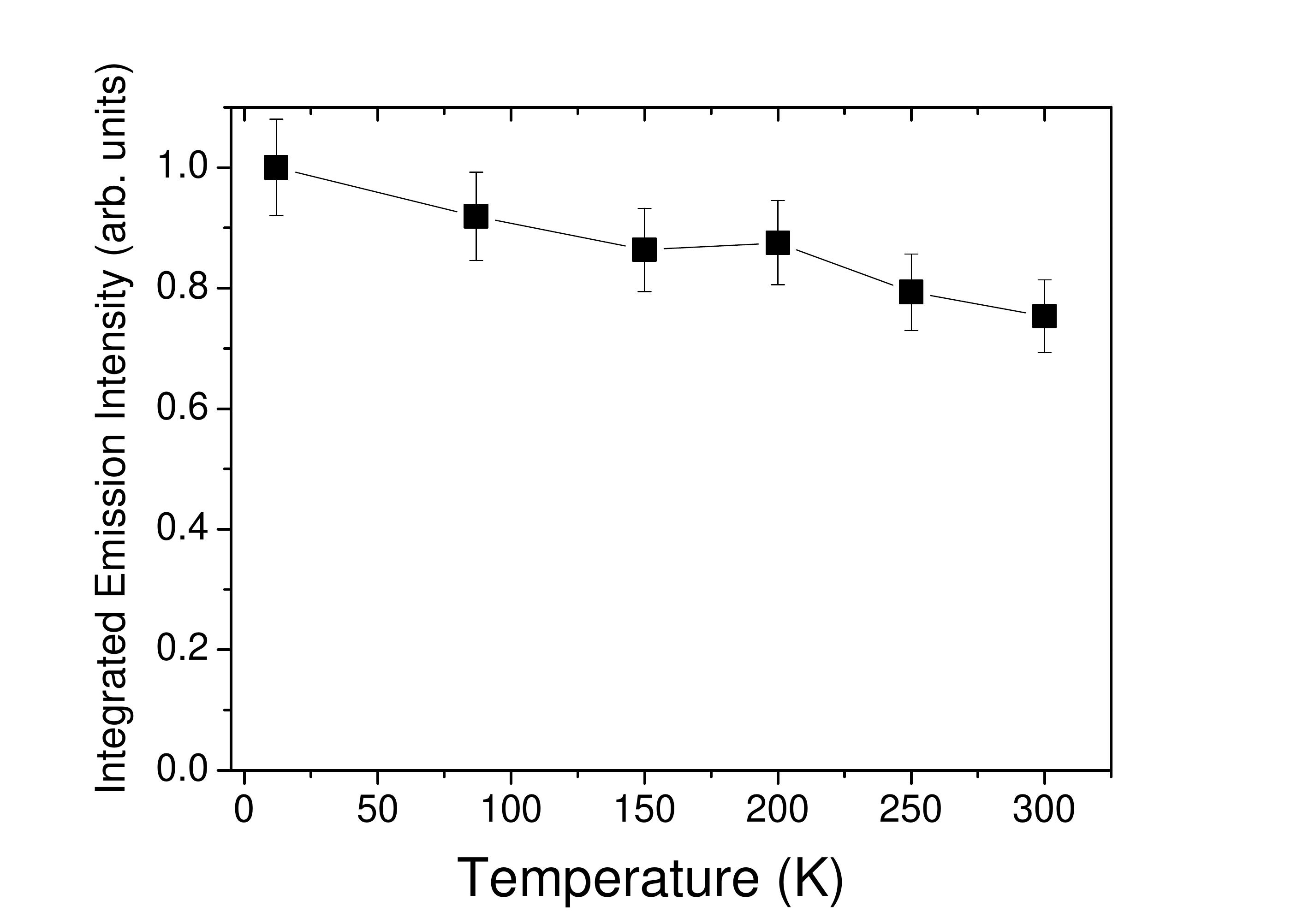}
\caption{{\scriptsize \sf Integrated relative photoluminescence intensity as a function of temperature, for the TPB sample evaporated on polymeric reflector substrate (layer density 175 $\mu$g/cm$^2$).} }
\label{fig:R7} 
\end{figure}
\begin{figure}[!htbp]
\centering
\includegraphics[width=10.cm]{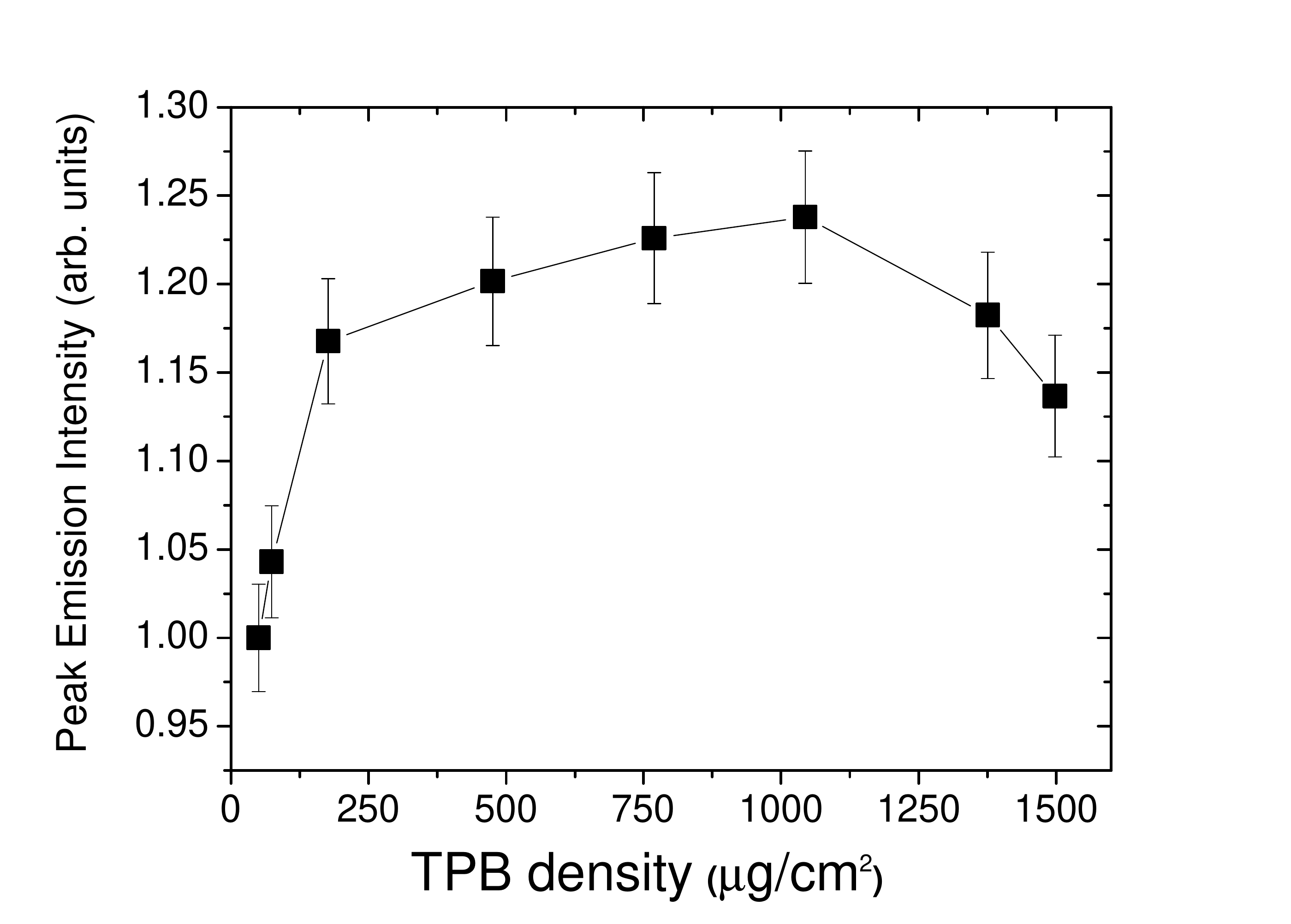}
\caption{{\scriptsize \sf RT photoluminescence relative peak intensity of TPB samples evaporated on polymeric reflector substrate, as function of the TPB layer density.} }
\label{fig:R8} 
\end{figure}
The observed vibronic replicas (corresponding, in order of increasing energy, to transitions involving a change of the vibrational quantum number of 1, 2, 3 and 4 in our case) are spaced in energy by 170 meV (1400 cm$^{-1}$) which closely corresponds to an effective vibrational mode associated to the stretching mode vibration of trans-butadiene \cite{rob-1,rob-2}. The vibrational structures are more pronounced at 87 K which is the liquid Argon temperature, at which the films are intended to be operated as wavelength shifters.
For the same sample we report in Fig.\ref{fig:R7}  the variation of the relative integrated PL intensity as a function of temperature, showing an increase of about 10\% of the luminescence in going from 300 K to 87~K.\\
The relative luminescence yield, measured at RT, of the TPB films evaporated on polymeric reflector substrate is plotted versus the film density in Fig.\ref{fig:R8}. The PL intensity grows starting from the lower densities, as more active centers (TPB molecules) become available to excitation, reaching a plateau starting around  200 $\mu$g/cm$^2$, with a slight increase up to 1000 $\mu$g/cm$^2$, after which the yield starts to decrease.

\subsection{TPB in polystyrene on glass substrate}
\label{sec:TPB-PS-glass}
TPB samples at different concentrations in PS matrix on glass substrate have been characterized in the UV - VUV range as a function of temperature.

\begin{figure}[!htbp]
\centering
\includegraphics[width=9.0cm]{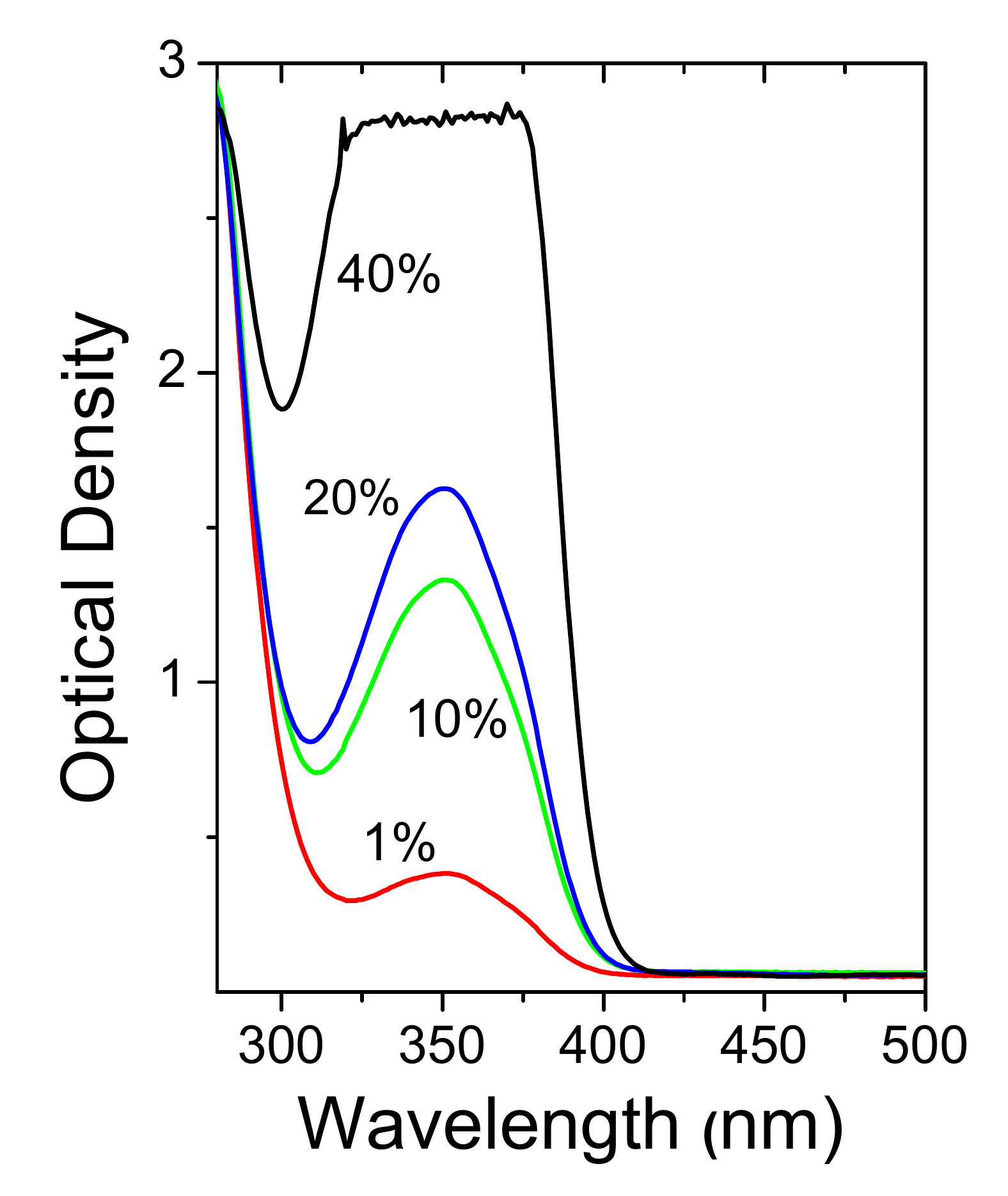}
\caption{{\scriptsize \sf RT absorption spectra of TPB in polystyrene on glass substrate, recorded at four different TPB concentrations 
(expressed as weight fraction of the solid solution).} }
\label{fig:R9} 
\end{figure}

The absorption spectra of TPB molecules dispersed in polystyrene are shown in Fig.\ref{fig:R9} for the four investigated TPB concentrations. 
An absorption band is observed, peaked  at about 350 nm. It corresponds to the HOMO-LUMO transition\footnote{Chemico-Physical properties of molecules can be interpreted by using the molecular orbital model. In particular the energy difference between the HOMO (Highest Occupied Molecular Orbital) and the LUMO (Lowest Unoccupied Molecular Orbital) corresponds to the lowest excitation energy.} of TPB \cite{rob-3,rob-4} and is responsible for the re-absorption mechanism described above. The absorption intensity should in principle be linearly dependent on the concentration of the absorbing centers for a given thickness of the film, and the observed deviation from linearity should be ascribed to the difficulty in controlling the thickness of the polystyrene film in the dip-coating process. The peak for the 40\% sample is not observed due to saturation of the spectrophotometer. 
\begin{figure}[!htbp]
\centering
\includegraphics[width=8.0cm]{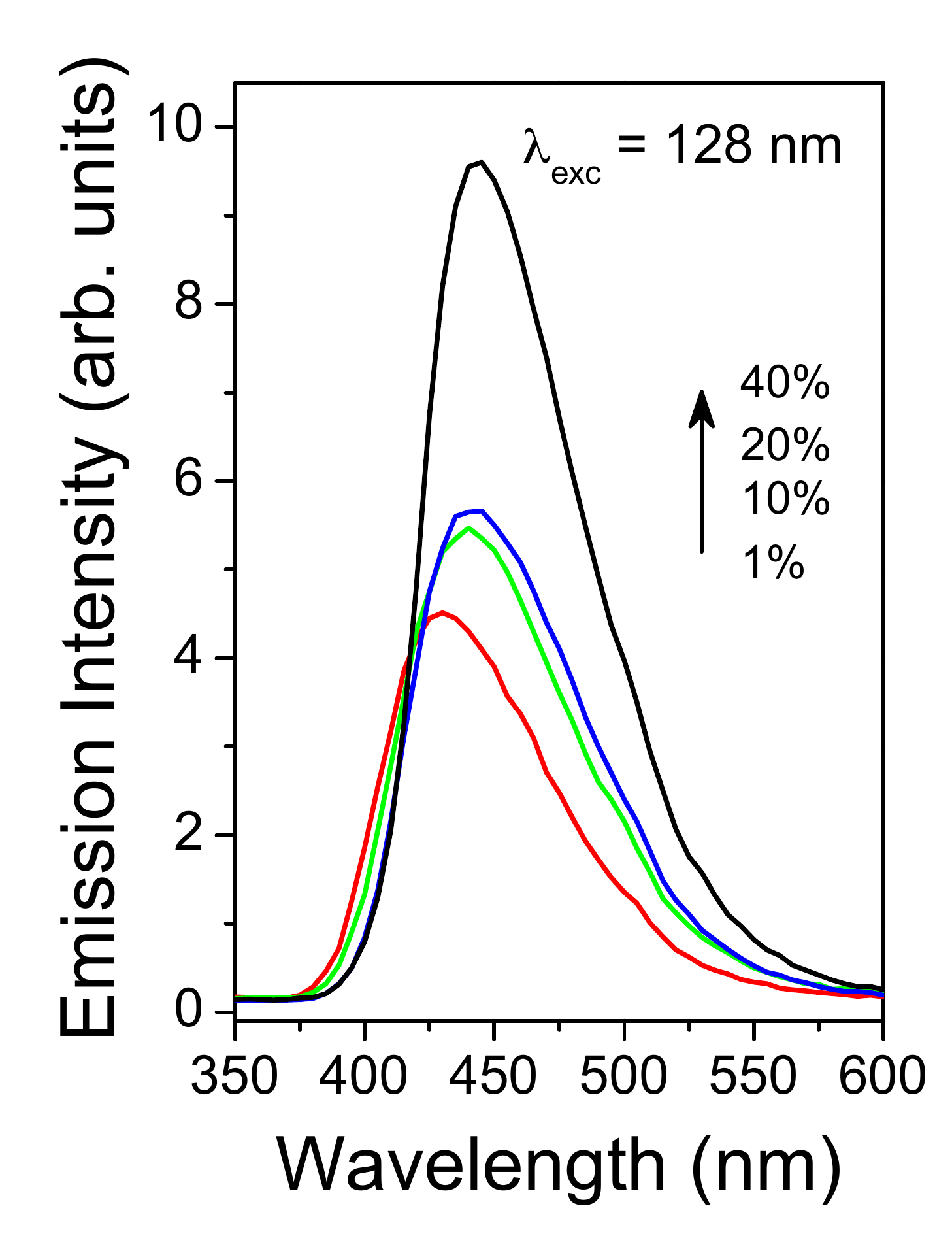}
\caption{{\scriptsize \sf RT photoluminescence spectra of TPB in polystyrene on glass substrate at various TPB concentrations (expressed as weight fraction of the solid solution). Photoluminescence spectra are excited at 128 nm.} }
\label{fig:R10} 
\end{figure}
\begin{figure}[!htbp]
\centering
\includegraphics[width=11.0cm]{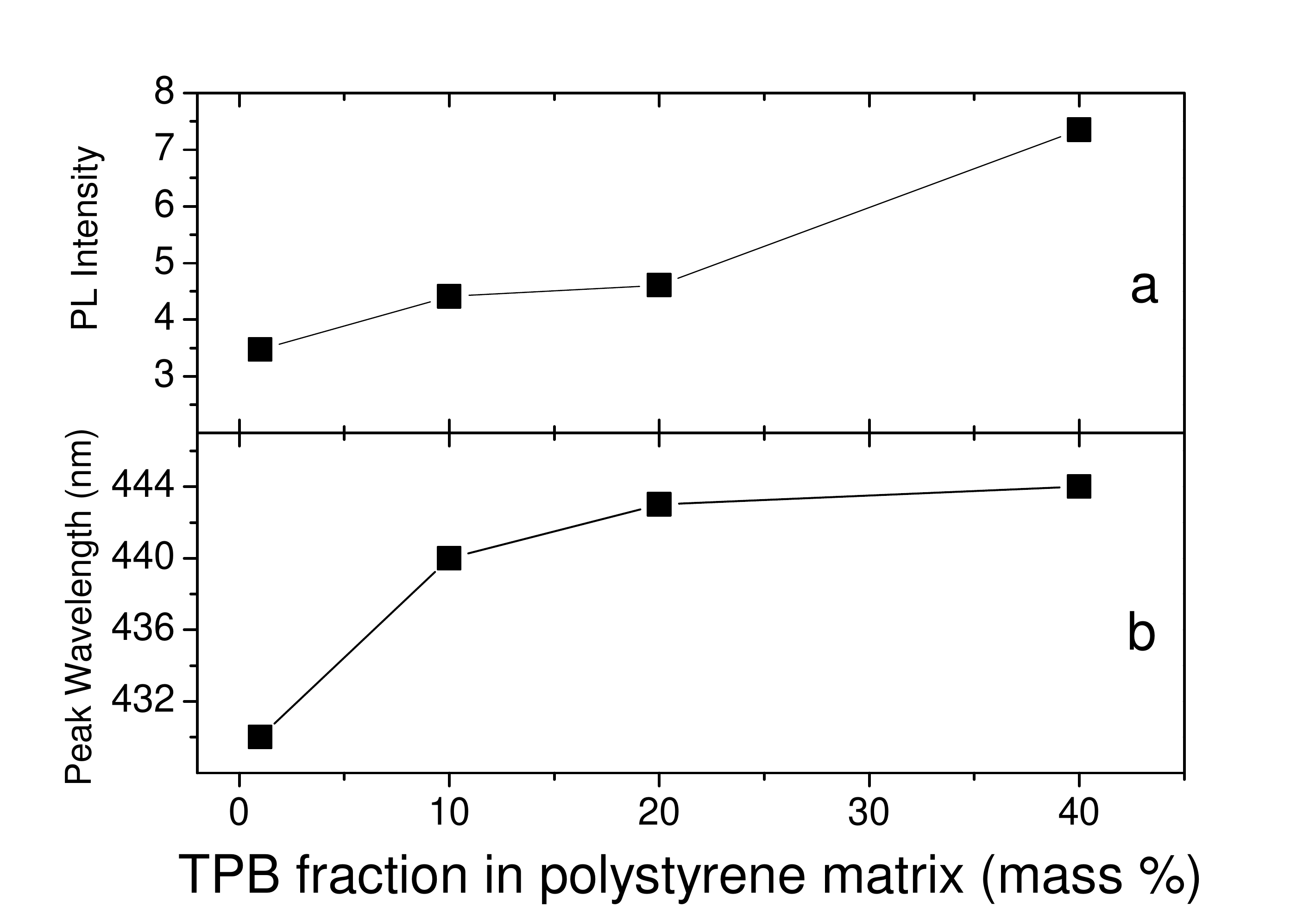}
\caption{{\scriptsize \sf Integrated relative photoluminescence intensities (a) and peak wavelength position (b) as a function of the TPB concentration (weight fraction \%) in polystyrene on glass substrate.} }
\label{fig:R11} 
\end{figure}
\begin{figure}[!htbp]
\centering
\includegraphics[width=11.0cm]{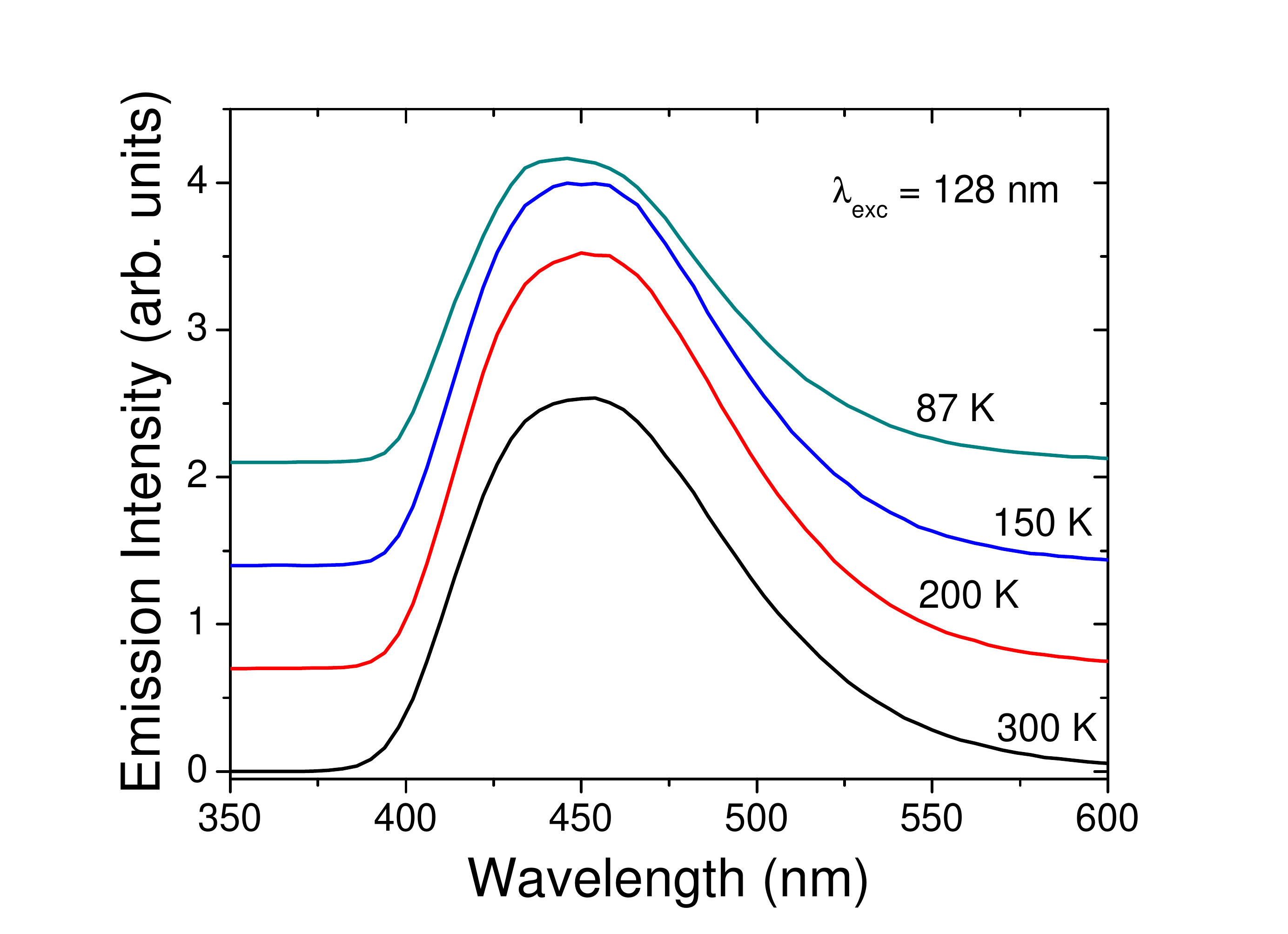}
\caption{{\scriptsize \sf Photoluminescence spectra of TPB in polystyrene on glass substrate (1\% sample) excited at 128 nm, as a function of temperature. } }
\label{fig:R12} 
\end{figure}
\begin{figure}[!htbp]
\centering
\includegraphics[width=7.5cm]{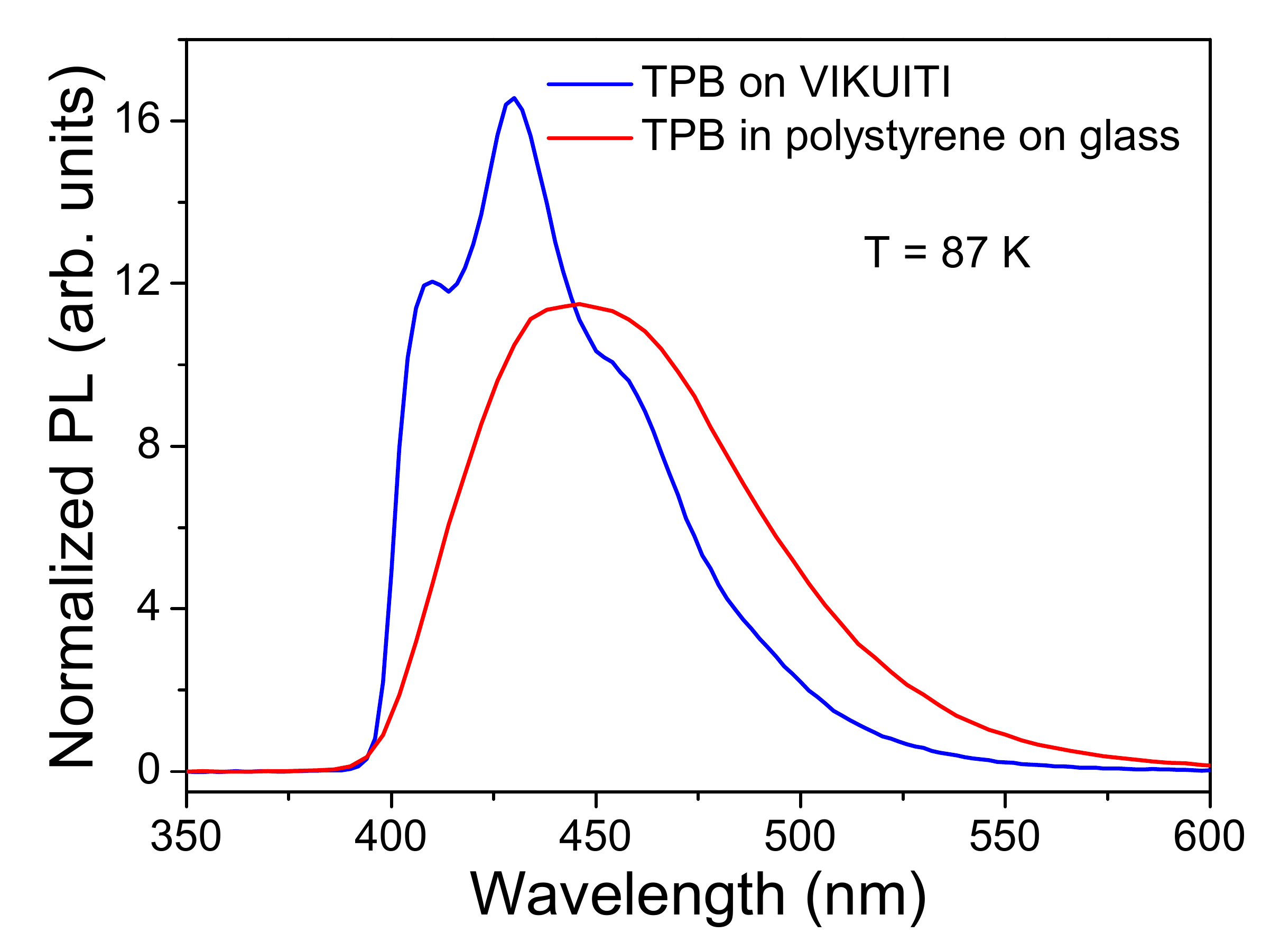}
\includegraphics[width=7.5cm]{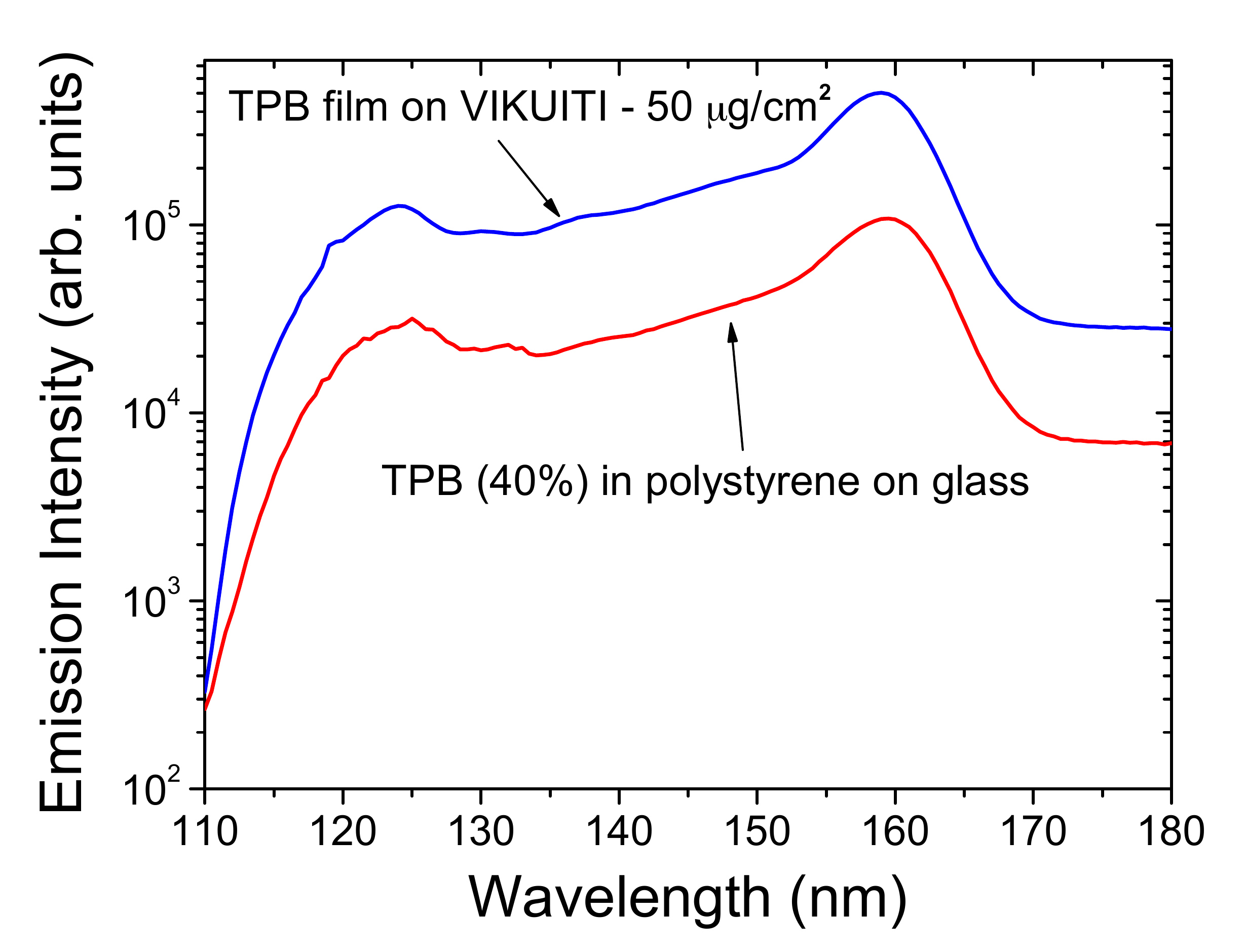}
\caption{{\scriptsize \sf [Left] Photoluminescence spectra of the 1\% sample of TPB in polystyrene on glass (red line) and of TPB evaporated on specular reflector substrate (layer density 175 $\mu$g/cm$^2$) (blue line), both excited at 128 nm and at liquid argon temperature  (87 K). The spectra are normalized to unit area. [Right] RT photoexcitation spectra in the VUV region of TPB film on reflector (layer density 50 $\mu$g/cm$^2$) and of TPB in polystyrene on glass (40\% sample). Both spectra are not corrected for the deuterium lamp spectral distribution. Note the vertical logarithmic scale. } }
\label{fig:R13} 
\end{figure}

By exciting at RT the same samples at 128 nm, we obtain the PL spectra shown in Fig.\ref{fig:R10}. The TPB emission bands show an overall shift toward lower energies when compared to the TPB emission of samples evaporated on polymeric substrates. 
Moreover, in going from 1\%  to 40\% concentration, a further shift to the red of the emission peak can be observed as shown in Fig.\ref{fig:R11}(b). \\
Again, re-absorption processes are responsible for the latter red-shift, which is clearly observed for the nominal 10\% and 20\%  TPB concentrations compared to the 1\% concentration. The re-absorption mechanism seems to be less efficient for the 40\% concentration for which the peak position is very close to that of the 20\% sample, but one has to take into account the fact that at higher concentrations the exciting light penetration into the sample is strongly reduced as well as the volume available for re-absorption.

The PL spectra of the 1\% sample as a function of temperature are shown in Fig.\ref{fig:R12}. A remarkable difference with pure TPB evaporated on 
glass or polymeric layers is that we do not observe the vibronic structures neither at room temperature nor at lower temperatures, and the emission lineshape is hardly affected by temperature. This behavior closely resembles that of TPB in liquid solutions, where the randomness of the environment geometry around the TPB molecule induces the observed inhomogenous broadening of the vibronic structures, which merge in a single, unresolved band \cite{rob-5,rob-6}.

Figure\ref{fig:R13} [Left] shows the PL spectrum of the 1\% sample of TPB in polystyrene on glass together with the PL spectrum of TPB evaporated on polymeric reflector substrate (layer density 175 $\mu$g/cm$^2$) both excited at 128 nm and at liquid argon temperature. The two spectra are normalized to unit area so that one can clearly appreciate the different spectral distribution of the emitted photons among the two samples. This fact can be of some relevance when one has to match the TPB emission with the photomultiplier spectral response in the liquid argon detectors.
Figure\ref{fig:R13} [Right]  shows the comparison of  the RT PE spectra in the VUV region for  the 40\% sample of TPB in polystyrene and the 50 $\mu$g/cm$^2$ sample of TPB evaporated on reflector substrate, monitored at the emission peak wavelength. The spectra are not corrected for the deuterium lamp spectral distribution, which is responsible for the overall shape of the spectra. One can notice how the wavelength shifting efficiency at 128 nm of the evaporated film is about 5 times higher that of TPB in polystyrene at the specified concentrations.

\section{Optical Transmittance and Reflectance}
\label{sec:Trans_Refl}
Hemispheric optical transmittance and reflectance were measured with a spectrophotometer for samples of TPB evaporated on specular reflector substrate and of TPB in polystirene matrix deposited on glass substrate.
The optical characteristics of the two substrates are very different (high reflectivity for the multi-layer plastic mirror substrate and high transmission for the glass) and are expected to play a major role in the transmittance and reflectance measurements of the different samples. 
 However, TPB is a very efficient photoluminescent material, and therefore ``spurious" contributions to the detected light are expected from TPB emission, even for thin film samples. A measure-correction procedure had to be designed and applied to eliminate these contributions. 
 
 As a matter of fact, while in a spectrophotometer the beam of light incident onto the sample is generally wavelength filtered, usually there is no additional spectral filter between sample and detector. Any PL excited by the impinging beam is therefore erroneously recorded by the instrument as part of the light which is transmitted or reflected - depending on the type of measurement being performed - by the sample. This fact, which practically translates into measures that are artificially higher than normal and even larger then unity within the photoexcitation band of the sample material, represents a severe source of systematic errors. This is most pronounced for the case of hemispheric measurements because in such a wide-angle setup photoluminescence is almost entirely collected by the system. 

We designed and applied a measurement procedure to eliminate the photoluminescence-induced errors and get corrected transmittance and reflectance spectra. 

The procedure, described in detail in \cite{ENEA-rep} (including a full treatment of error propagation) implies the use of a suitable low-pass optical filter and multiple measurements which need to be post-processed. Restricting for conciseness the discussion to the case of optical transmittance (similar considerations hold for reflectance), the adopted starting hypothesis is that photoluminescence perturbs the measure according to
\begin{equation}
T_{pert}(\lambda)~=~T(\lambda)~+~\frac{\Omega}{4\pi}~N~\sigma_{PE}(\lambda)\int_{\Lambda_{PL}}\Gamma(\lambda')~\sigma_{PL}(\lambda')~d\lambda'
\end{equation}
In the previous equation, $\lambda$ is the wavelength,  $T(\lambda)$ and $T_{pert}(\lambda)$ are respectively the real and perturbed spectral transmittances of the sample, $\Omega$  is the acceptance solid angle of the detector,  $N$ is the number of excited photoluminescent centres in the sample during the measurement, $\sigma_{PE}(\lambda)$  and $\sigma_{PL}(\lambda)$ are the wavelength-dependent cross sections for photoexcitation and photoluminescence respectively, and  $\Gamma(\lambda)$
is an instrumental parameter that represents the global spectral response and conversion electronics of the spectrophotometer - it essentially depends on the instrument setup.  The integral is performed over the wavelength domain $\Lambda_{PL}$ where the photoluminescence cross section is non-zero. The above equation represents the fact that the measured results are perturbed only at those wavelengths that fall within the domain of non-zero photoexcitation cross section; elsewhere it can be assumed to be error-free, as far as this kind of error is considered.
After having recorded the spectrum $T_{pert}(\lambda)$, one can record another measurement with the insertion of a suitable low-pass filter.
\begin{figure}[!htbp]
\centering
\includegraphics[width=9.0cm]{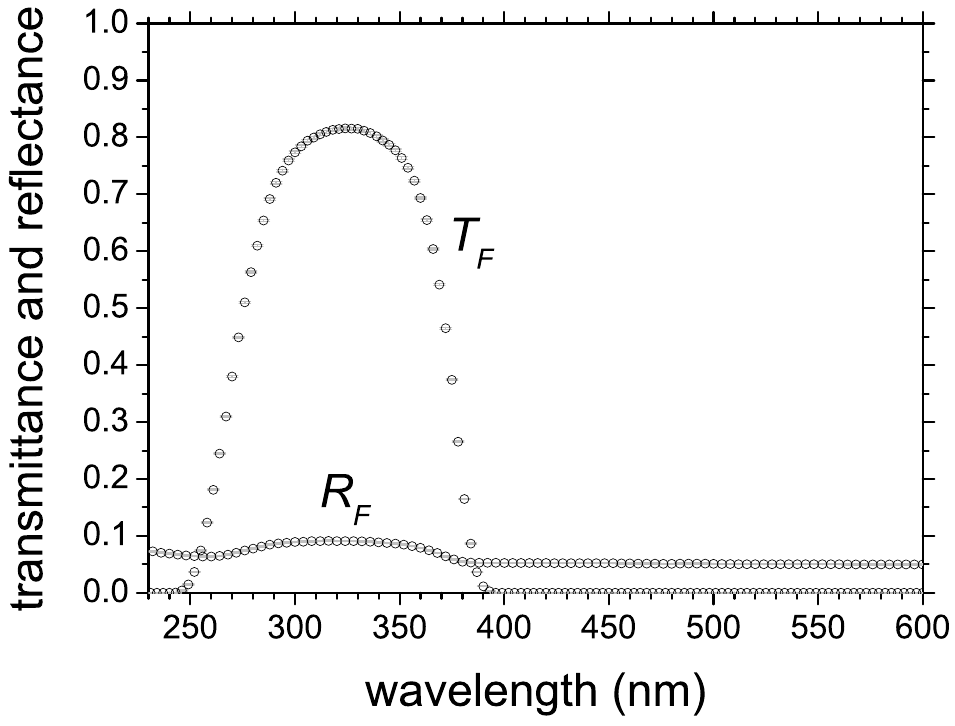}
\caption{{\scriptsize \sf Optical transmittance and reflectance,  $T_F(\lambda)$ and $R_F(\lambda)$, of the low pass filter selected to correct the measured photometric spectra according to our correction procedure. } }
\label{fig:E1} 
\end{figure}
 The optical transmittance $T_{F}(\lambda)$ of such a filter should ideally be  as small and constant as possible, $T_{F}(\lambda)\approx \epsilon \ll 1$, at the wavelengths in $\Lambda_{PL}$, much larger outside $\Lambda_{PL}$,  $T_{F}(\lambda)\gg \epsilon$. By performing a transmittance measurement with such a filter placed between sample and detector, the recorded spectrum should be
\begin{equation}
T^{filter}_{pert}(\lambda)~=~T_F(\lambda) T(\lambda)~+~\frac{\Omega}{4\pi}~N~\sigma_{PE}(\lambda)\int_{\Lambda_{PL}}\Gamma(\lambda')~T_F(\lambda')~\sigma_{PL}(\lambda')~d\lambda'
\end{equation}
Thanks to the required conditions for the filter, can be approximated as
\begin{equation}
T^{filter}_{pert}(\lambda)~\approx~T_F(\lambda) T(\lambda)~+~\epsilon~\frac{\Omega}{4\pi}~N~\sigma_{PE}(\lambda)\int_{\Lambda_{PL}}\Gamma(\lambda')~\sigma_{PL}(\lambda')~d\lambda'
\end{equation}
From the two last equations, the true transmittance of the sample in the wavelength region of photoemission can be extracted as
\begin{equation}
T(\lambda)~\approx~\frac{T^{filter}_{pert}(\lambda)~-~\epsilon~T_{pert}(\lambda)}{T_F(\lambda)~-~\epsilon}
\label{eq:Tlambda}
\end{equation}
It should be noticed that the previous equation cannot be always applied at those photoemission wavelengths for which $T_F(\lambda)\approx \epsilon$  because, taking uncertainties into account, the estimated uncertainty associated to the derived real transmittance could become meaninglessly larger than one \cite{ENEA-rep}; minimizing such a wavelength region inside which our procedure cannot be applied is hence a further criterion for selecting the most suitable optical filter among the available ones. In our case, this possibly troublesome wavelength interval corresponds to about 380-410 nm. The spectral characteristics of the selected filter are plotted in Fig.\ref{fig:E1}.

As previously mentioned, a procedure similar to the one so far illustrated for the transmittance holds also for the correction of the experimental reflectance spectra. Let us report here just the main result, i.e. the equation, equivalent to Eq.(\ref{eq:Tlambda}), which should be used to get the real reflectance spectrum from the two perturbed ones (one with the filter interposed and the other without it),
\begin{equation}
R(\lambda)~\approx~\frac{R^{filter}_{pert}(\lambda)~-~R_F(\lambda)~-~\epsilon~T_F(\lambda) R_{pert}(\lambda)}{T_F(\lambda)~[T_F(\lambda)~-~\epsilon]}
\label{eq:Rlambda}
\end{equation}
with obvious meaning of the used symbols. Note that Eq.(\ref{eq:Rlambda}) looks slightly more complex than Eq.(\ref{eq:Tlambda}) because the low-pass filter not only filters the light beam which is reflected by the sample, but also the one impinging onto it. Details about the utilizeded optical filter, further information concerning the correction of spectra perturbed by induced photoluminescence of the sample, and determination of measurement-uncertainty propagation for the application of the procedure are fully reported and discussed in \cite{ENEA-rep}.

\subsection{Experimental set-up}
\label{sec:exp_set-up-bis}
The hemispheric optical transmittance and reflectance of the samples were measured at RT in the spectral range 250-600 nm by means of a Perkin-Elmer Lambda-19$\textsuperscript{\textregistered}$ spectrophotometer equipped with an integrating sphere of 150 mm diameter.
The Lambda-19$\textsuperscript{\textregistered}$ operates in dual-beam mode, meaning that the specimen is placed in the path of the sample beam, while a second beam provides a reference for correcting lamp-intensity fluctuations. Background signals are automatically filtered out by the instrument via synchronous detection of light. For the reflectance measurements, suitable and calibrated reference mirrors and BaSO$_4$ capsules were used to record the so-called ``baselines". The baselines were then used to normalize the specimen spectra in order to derive their absolute reflectance.

\subsection{Reflectance of TPB evaporated on specular reflector substrate}
\begin{figure}[!htbp]
\centering
\includegraphics[width=8.0cm]{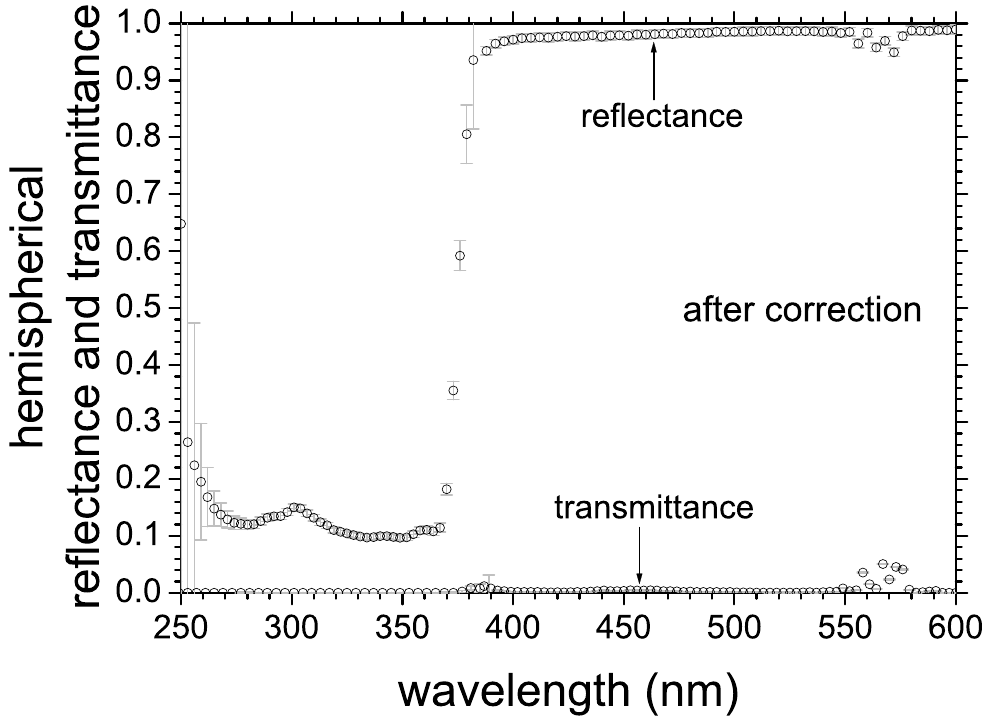}
\caption{{\scriptsize \sf  Reflectance and transmittance spectra of the specular reflector substrate. The filter procedure described in 
Sec.4 has been applied to eliminate perturbations due to (residual) photoluminescence of the substrate itself.} }
\label{fig:RT-VM2000} 
\end{figure}

The hemispherical reflectance of a surface layer of TPB deposited over specular reflector substrate 
is important for DM detectors characterization. 
The reflectance (and transmittance) of the bare specular reflector substrate has been first measured (see Fig.\ref{fig:RT-VM2000}) and found consistent with 
available data from the manufacturer \cite{vikuiti}.

A sample with TPB density of 370  $\mu$g/cm$^2$ over VM2000\texttrademark~ foil
was measured in the wavelength range 250-600 nm, with the TPB side facing the incident light beam of the spectrophotometer. 
The photoexcitation spectrum  of TPB deposited on reflector substrate, shown in Fig.\ref{fig:R5}, represents the wavelength interval where the measured reflectance is very likely perturbed by photoluminescence and needs to be corrected with the explained procedure - see Eq.(\ref{eq:Rlambda}). 
\begin{figure}[!htbp]
\centering
\includegraphics[width=7.5cm]{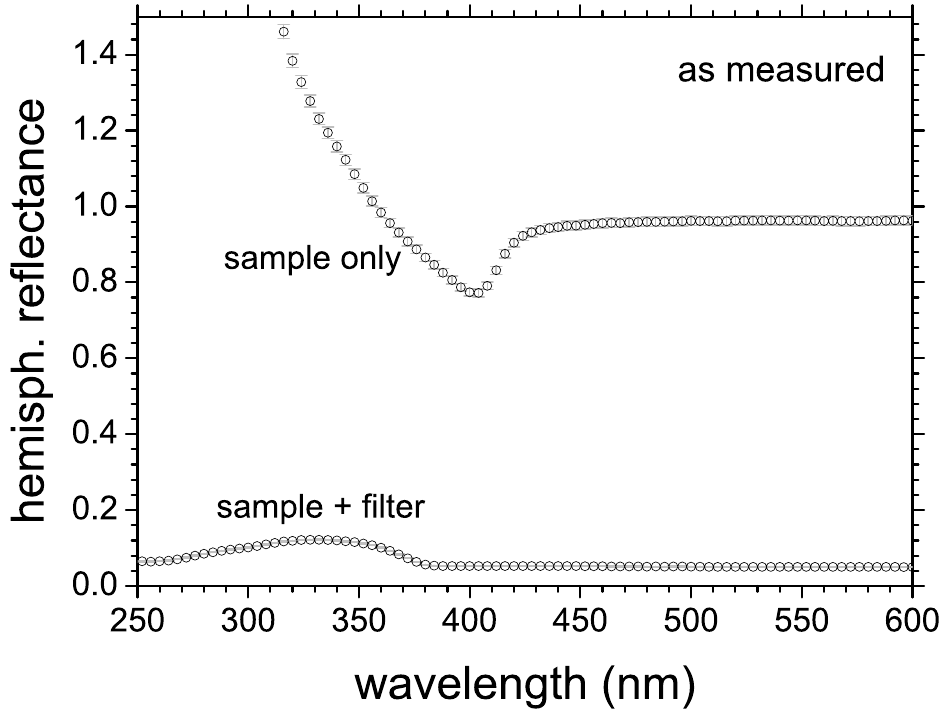}
\includegraphics[width=7.5cm]{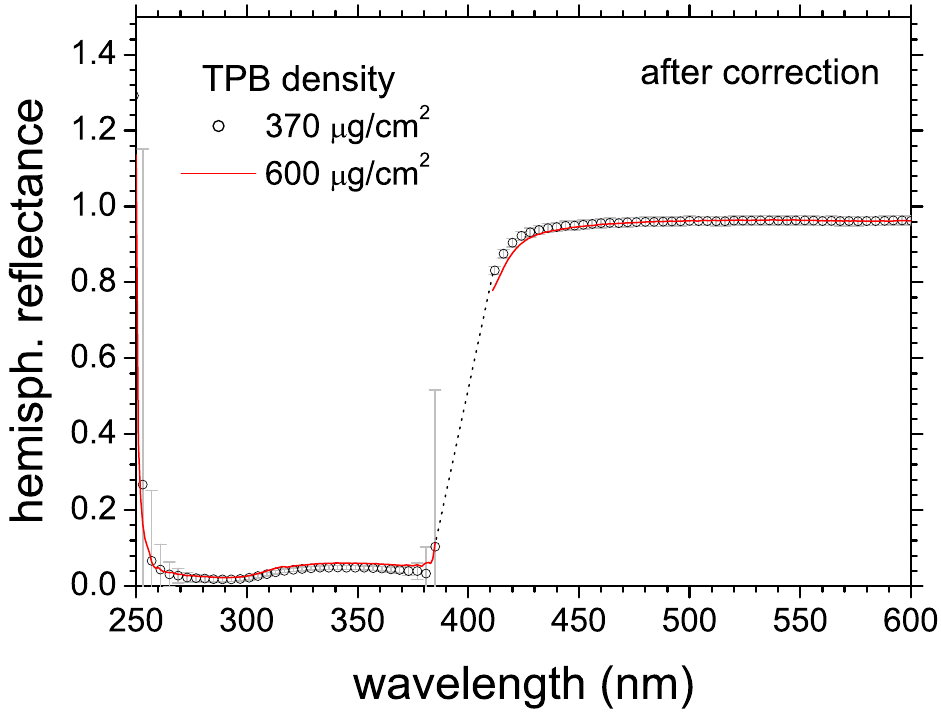}
\caption{{\scriptsize \sf Hemispherical reflectance spectra of a TPB surface layer deposited (density of 370  $\mu$g/cm$^2$) over a polymeric reflector substrate: as measured with and without low-pass filter [Left], and after application of the procedure to eliminate perturbations due to photoluminescence [Right]. Also shown (red curve) is the result from a reflector sample with higher TPB deposit density (600  $\mu$g/cm$^2$).} }
\label{fig:E2} 
\end{figure}

Therefore, the measurement of the hemispherical reflectance was repeated according to our procedure after the insertion of the selected low-pass filter (see filter reflectance and transmittance spectra in Fig.\ref{fig:E1}) and the real hemispherical reflectance of the sample was estimated by applying Eq.(\ref{eq:Rlambda}) to the recorded spectra. The ``as measured" and corrected spectra are shown
 in Fig.\ref{fig:E2}.\\
The corrected reflectance spectrum of Fig.\ref{fig:E2} [Right] supports high reflectance, larger than 90\% at wavelengths above 410 nm, and around 95\% at 430 nm, at the onset and peak of the TPB photoluminescence spectrum (Fig.\ref{fig:R6}), and it was verified with other samples that  to first approximation it is independent of the TPB thickness in the 200 to 600 $\mu$g/cm$^2$ range.

This is a noticeable result that fully justifies the choice of the polymeric specular reflector with TPB coating for the boundary surfaces of LAr detectors, where the highest wavelength-shifting efficiency for VUV photons (from LAr scintillation) and the maximum reflectivity in the UV-Vis range (from TPB emission) should be simultaneously established.

\subsection{Transmittance and reflectance of TPB in polystyrene on glass substrate}
\begin{figure}[!htbp]
\centering
\includegraphics[width=8.0cm]{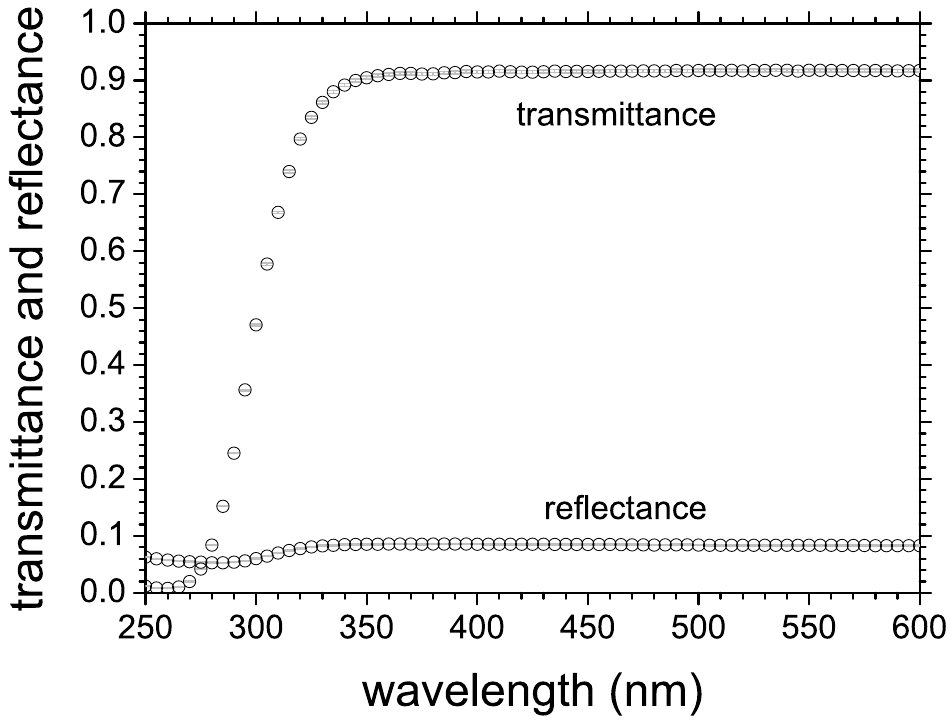}
\caption{{\scriptsize \sf Transmittance and reflectance spectra of a 1 mm thin glass substrate of the same kind as that over which a polystyrene-diluted-TPB layer was deposited. } }
\label{fig:E3} 
\end{figure}

Another type of sample, consisting of a thin layer of TPB in polystyrene matrix (20\% weight fraction) over a 1 mm thick glass slab, was optically characterized in the same wavelength range from 250 nm to 600 nm. In this case, the transparency of the glass substrate in a large extent of the examined spectral range
allowed the measurement of hemispherical optical transmittance of the sample in addition to reflectance. Having optically analyzed a bare glass substrate of the same kind - see Fig.\ref{fig:E3}  -  some features of the polystyrene-diluted-TPB layer itself could be inferred from the available data.

\begin{figure}[!htbp]
\centering
\includegraphics[width=7.5cm]{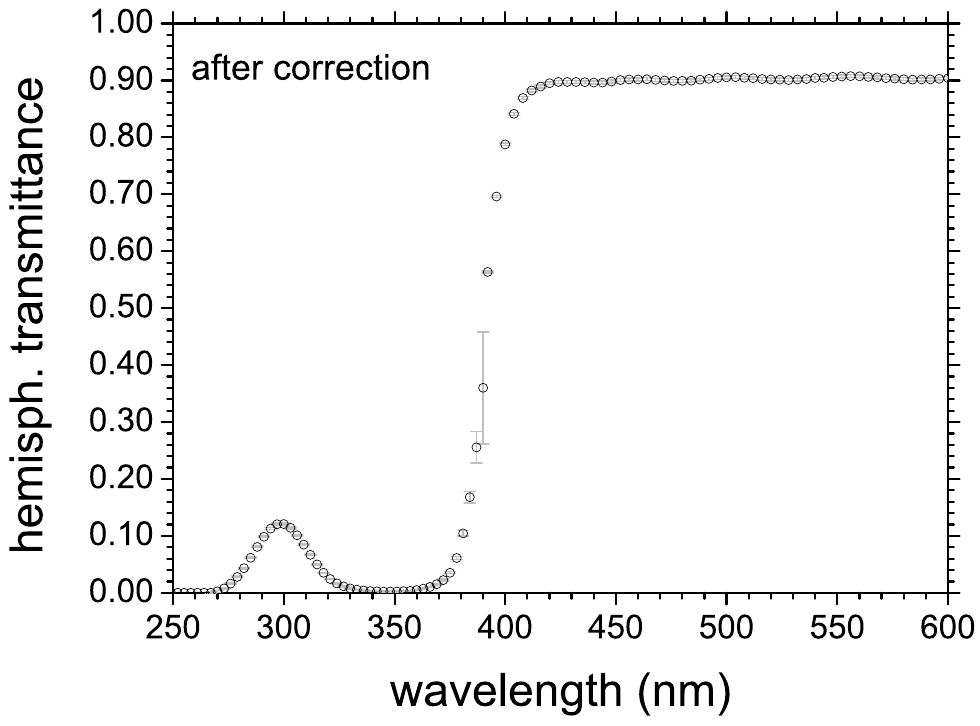}
\includegraphics[width=7.5cm]{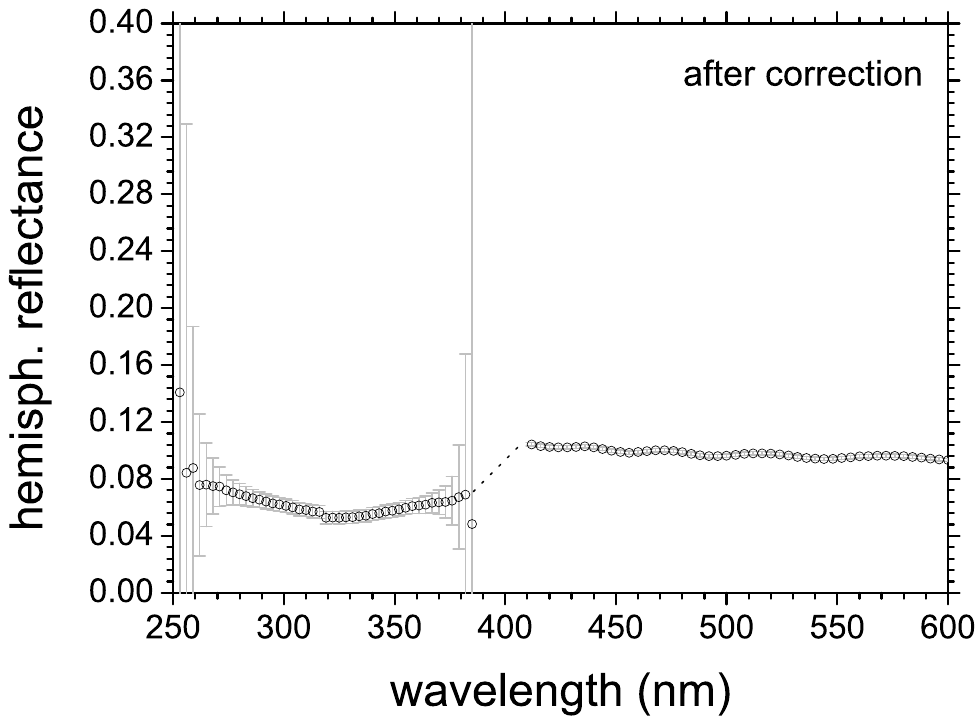}
\caption{{\scriptsize \sf Hemispherical transmittance [Left] and reflectance [Right] spectra of a polystyrene-diluted-TPB layer (20\% weight fraction) deposited over a 1 mm thin glass substrate, corrected from unwanted photoluminescence according to our correction procedure.} }
\label{fig:E4} 
\end{figure}
As for the pure-TPB layer evaporated over polymeric reflector considered in the previous section, also in this case the photoluminescent nature of TPB gave rise to systematic errors in the spectra measured by means of the Lambda 19 spectrophotometer. Therefore, the correction procedure described above was successfully applied through the elaboration of measurements taken with and without the low-pass filter of Fig.\ref{fig:E1}. The resulting  spectra are shown in Fig.\ref{fig:E4}.
Transmittance is at 90\% level and reflectance at 10\% for all the wavelengths above 400 nm.\\
One can notice that transmittance, Fig.\ref{fig:E4} [Left], features two minima at about 250 nm and 350 nm. A comparison with the corresponding reflectance spectrum - Fig.\ref{fig:E4} [Right] -  demonstrates that such minima correspond to optical absorption of the sample, either by the TPB+polystirene film or by the substrate (or both). In particular, the spectra of Fig.\ref{fig:E3} suggest that while the feature at 250 nm is mainly due to the substrate glass, at wavelengths longer than $\sim$300 nm the substrate does not absorb light, so that the minimum of transmittance at about 350 nm should be attributed to the polystyrene-diluted-TPB layer alone. 
As already pointed out in Sec.\ref{sec:TPB-PS-glass}, such an absorption band corresponds to the HOMO-LUMO transition of TPB.
  As far as reflectance is concerned, Fig.\ref{fig:E4} [Right] demonstrates that, besides a quite weak interferential effect visible from 400 nm to 600 nm, the reflectance of the sample is closely comparable with that of the substrate alone, so that one can deduce that the optical constants of glass and layer should be almost matching with each other.

\section{Conclusion}
The main spectral features and optical characteristics of TPB films on different substrates, commonly employed elements of DM detector optical systems, have been investigated and presented here to provide means for more accurate simulations and designs of current and future LAr experiments.\\
 TPB emission spectra by VUV excitation at 128 nm with lineshape and relative intensity variation down to LAr temperature were measured. For TPB films evaporated on specular reflector substrates, the unresolved vibronic structures at room temperature become clearly visible as the temperature is decreased. At 87 K one can observe at least four distinct structures and the PL intensity  increases of about 10\% with respect to the luminescence at room temperature.
A remarkable difference with TPB diluted in polymeric layers on glass is that we do not observe the vibronic structures neither at room temperature nor at lower temperatures, and the emission lineshape is hardly affected by temperature. This behavior closely resembles that of TPB in liquid solutions, where the randomness of the environment geometry around the TPB molecule induces the observed inhomogenous broadening of the vibronic structures, which merge in a single, unresolved band.

Hemispherical reflectance and transmittance of the TPB coated substrates in the blue wavelength range of the TPB emission spectrum are fundamental detector parameters.
Reflectance of the (boundary) mirror surfaces after TPB film deposition and transmittance of (PMT window) glass with TPB-polystyrene coating were measured. 
The TPB-coated mirror surface spectrum, after careful correction of the perturbation due to TPB photoluminescence, supports high reflectance, larger than 90\% at wavelengths above 410 nm, and around 95\% at 430 nm, at the onset and peak of the TPB photoluminescence spectrum, and it is in first approximation independent of the TPB thickness in the 200 to 600 $\mu$g/cm$^2$ range.
This result fully supports the choice of the specular reflector with TPB coating for the boundary surfaces of LAr detectors.
Transmittance of TPB in polystyrene on glass substrate, after perturbation correction, is found at 90\% level and reflectance at 10\% for all wavelengths above 400 nm. This result justifies this type of solution currently adopted by some experiments.

\section*{Aknowlegdments}
We recognize the importance of fruitful discussions and the role of the stimulating environment
of the WArP experiment, which definitively triggered and motivated the effort that led to the results presented in this paper.
In particular, we thank INFN Pavia for the vacuum evaporator used for the samples preparation with TPB deposition.
We acknowledge the financial support of MIUR PRIN - 2007WKA8F8-002 (2007).


\end{document}